\documentclass[review]{elsarticle}

\usepackage{hyperref}
\usepackage{color}
\usepackage{ifthen}
\usepackage{soulutf8} 
\usepackage{multirow}
\usepackage{longtable}
\usepackage{graphicx}
\usepackage{tabularx}
\usepackage{float}
\usepackage{flushend}
\usepackage{todonotes}
\usepackage{url}
\usepackage[fleqn]{amsmath}
\usepackage{caption}
\usepackage{placeins}
\usepackage{lscape}
\usepackage{booktabs}
\usepackage{listings}

\journal{Journal}

\begin{document}
\begin{frontmatter}

\title{Consistent and Compatible Modelling of Cyber Intrusions and Incident Response Demonstrated in the Context of Malware Attacks on Critical Infrastructure}



\author[1]{Peter Maynard}
\author[1]{Yulia Cherdantseva}
\author[2]{Avi Shaked}
\author[1]{Pete Burnap}
\author[1]{Arif Mehmood}

\address[1]{School of Computer Science and Informatics, Cardiff University, UK (e-mail: \{MaynardPG, CherdantsevaYV, BurnapP, MehmoodA3\}@cardiff.ac.uk)}
\address[2]{Department of Computer Science, University of Oxford, UK (e-mail: avishakedse@gmail.com)}





\begin{abstract}
Cyber Security Incident Response (IR) Playbooks are used to capture the steps required to recover from a cyber intrusion. Individual IR playbooks should focus on a specific type of incident and be aligned with the architecture of a system under attack. 
Intrusion modelling focuses on a specific potential cyber intrusion and is used to identify where and what countermeasures are needed, and the resulting intrusion models are expected to be used in effective IR, ideally by feeding IR Playbooks designs. IR playbooks and intrusion models, however, are created in isolation and at varying stages of the system's lifecycle.
We take nine critical national infrastructure intrusion models – expressed using Sequential AND Attack Trees – and transform them into models of the same format as IR playbooks. We use Security Modelling Framework for modelling attacks and playbooks, and for demonstrating the feasibility of the better integration between risk assessment and IR at the modelling level. 
This results in improved intrusion models and tighter coupling between IR playbooks and threat modelling which -- as we demonstrate -- yields novel insights into the analysis of attacks and response actions. The main contributions of this paper are (a) a novel way of representing attack trees using the Security Modelling Framework, (b) a new tool for converting Sequential AND attack trees into models compatible with playbooks, and (c) the examples of nine intrusion models represented using the Security Modelling Framework.

\end{abstract}

\begin{keyword}
Attack Trees, Critical National Infrastructure, Formal Modelling, Industrial Control Systems, Risk Assessment, Threat Modelling
\end{keyword}
\end{frontmatter}

\section{Introduction}

Advanced and persistent cyber security attacks on Critical National Infrastructure (CNI) present a significant risk for modern digital society. For example, an attack could result in an electric power cut in the case of an attack on the power grid \cite{nguyen2020electric}, in an environmental disaster in case of an attack on wastewater infrastructure \cite{hassanzadeh2020review}, or in a global disaster in case of an attack on nuclear facilities. The world was lucky enough to avoid the latter for now, but the Stuxnet attack offered a terrifying preview of the grim possibility if the impact of such attacks proliferate beyond self-destruction of centrifuges \cite{bakic202110}. 

To stay cyber resilient, CNI needs to invest in both minimising the chances of an attack succeeding and in minimising the consequences of an attack. In this paper, we discuss and demonstrate how these two tasks could be achieved via a consistent and compatible modelling of cyber threats and responses. In the conceptual sense, our work brings closer together cyber security risk management and incident response which are two intertwined areas currently lacking effective practical ways of feeding information from one into another \cite{ahmad2021can}. 

An organisation could minimise the chances of an attack succeeding via a better understanding of the threat landscape and effective threat modelling. Threat modelling techniques (e.g. Attack Trees \cite{schneier_1999} or STRIDE \cite{shostack_threat_2014}) are used to better understand how a system might fail under a cyber-attack. It forms part of the larger risk management lifecycle \cite{khalil_2024} and identifies potential threats to a system along with identifying potential countermeasures. Typically, this happens periodically throughout the entire lifecycle of the system, but primarily amid the design and development phase~\cite{shaked2023model}. Threat modelling is a mature field, often required to comply with existing cyber security standards \cite{cichonski_2012}. There are many well-researched methodologies and standards available, including for the Industria Control Systems (ICS) and Supervisory Control and Data Acquisition (SCADA) systems \cite{cherdantseva_2016}.

Incident Response (IR) is a process that is executed in the event of a cyber intrusion. IR defines a set of actions needed to contain the intrusion and to restore the system to a secure working state. The steps are normally captured in the form of a process model referred to as a playbook (or runbook). No two actionable 
playbooks are the same because no two cyber incidents are the same, and because each playbook depends on the specifics of system architecture and requires the interaction of people, technology, and processes which vary between organisations \cite{shaked_model_2022}. IR as a practice is less mature than threat modelling and is in the earlier stages of its development. While it is becoming more standardised, there is a lack of a consistent and interoperable approach for playbook modelling \cite{9557787}. Existing IR playbook formats such as CACAO, IACD, RECAST and RE\&CT \cite{9557787} are error-prone and verbose, resulting in playbooks which are inconsistent and incompatible with the broader risk management methodologies. The immaturity of playbook modelling formats further exacerbates the challenges of integration between risk management and IR. 

Threat modelling is a part of risk assessment process used to identify a weakness within a system, while IR defines a way to recover a system from an exploited weakness. While these processes should complement each other, mutually feeding information into each other, in practice they are often developed and implemented in isolation and independently of each other. This results in two distinct sets of modelling methods and tools, one for threat modelling and another for IR, with little to no cross-fertilisation. This leads to poor or ineffective coordination between risk management and IR, and to the loss of information and lessons learnt that could enrich and enhance the other process \cite{shaked_operations-informed_2023, ahmad2021can}. Furthermore, by performing threat modelling and IR modelling using different methods and languages, it is easier to overlook dependencies, direct and indirect links between events and possible process improvements as well as countermeasures for a threat vector.

The further evidence of the existing gap and the lack of coordination between risk assessment and incident response, could be found in \cite{schlette2023you}, where a detailed analysis of 1217 playbooks was conducted and risk had no representation in the analysed playbooks and is not discussed in the paper.  Additionally, in \cite{ferreira2023methodology}
a significant gap in assessing organisational responsiveness to cyber security attacks from a risk management perspective is acknowledged.

In this paper, we address the gap discussed above by proposing the use of a single consistent modelling approach for modelling both cyber-attacks and IR playbooks. In this work, we use Security Modelling Framework (SecMoF) \cite{shaked_model_2022,shaked_operations-informed_2023} for modelling attacks and playbooks, and for demonstrating the feasibility of the better integration between risk management and IR at the modelling level.  SecMoF \cite{secmof} is an Open-Source Eclipse-based modelling tool for the design and editing of Operations-Informed Incident Response Playbooks (OIIRP) ~\cite{shaked_operations-informed_2023}. 
SecMoF incorporates three modelling modules and supports the creation of: 
\begin{itemize}
    \item {Formalised Response to Incident Process Playbook (FRIPP) \cite{shaked_model_2022} representing the IR-focused approach by capturing IR as a concise process model};
    \item {Dependency Models \cite{slater_2016} representing a typical risk management-focused approach which  is based on constructing a data model for risk assessment in a positivistic top-down approach.}
    \item {OIIRP, which is an integration of FRIPPs and dependency models into a mutually dependent model offering a unified approach to capturing both IR steps and operational impact that IR steps incur in a system.}
\end{itemize}

We posit that the use of SecMoF -- with its novel approach to modelling OIIRP (as an integration of a playbook and dependency model) -- for modelling cyber-attacks enables the development of consistent and compatible IR playbooks and attack models, and yields novel insights into the analysis of attacks and response actions. In this paper, we demonstrate the novel use of SecMoF for modelling and analysing malware attacks on CNI. By taking Sequential AND (SAND) Attack Trees \cite{maynard_2020}, modelling them using SecMoF and comparing the SAND Attack Tree models with the OIIRP attack models, we highlight the advantages of using SecMoF for attack modelling and propose suggested improvements for both representations.

There are significant advantages that a consistent and compatible representation of cyber-attacks and playbooks using SecMoF offers to both risk management and IR practices. First, by using a unified modelling approach and creating compatible models, we are linking the two otherwise disconnected areas of risk management and IR allowing the learning and information from each practice to be shared between them more effectively, minimising omissions. 
For example, consider the case in which a new threat is identified and an attack model is developed. During this process, if an alternate notation is used then any points learnt or insights from this process could be lost because of the lack of an effective way of converting this new knowledge into the format acceptable by risk assessment methods. However, if the same modelling approach is used, then new insights, proposed actions and changes could be effectively and directly incorporated within IR playbooks and interconnected dependency models. 

Second, SecMoF enables the integration of an attack model with an operational model expressed via a dependency model, allowing for a detailed analysis of the impact of attack steps on system operations. In our previous work \cite{shaked_operations-informed_2023}, we demonstrated the advantages of combining a playbook with the operational model for analysing the impact that response actions could have on a system using dependency modelling \cite{slater_2016}. In this paper, we provide an example of the integration of an attack model with a dependency model and prove the feasibility and usefulness of this approach for attack models. 

Third, consistent modelling of attacks and playbooks provides better accessibility for non-technical operators as the Dependency Model is built up using information from domain experts, which typically IR Security Operations Centre (SOC) analysts may not be.  
Taking the same approach of lowering the complexity of understanding such actions, it is also possible for non-technical business experts to gain an understanding of intrusions into a system and for them to provide input to the threat modelling process.
This improves the organisational learning within CNI organisations as this information is retained in a concise, structured way, and technical information from playbooks is accompanied by / linked with information from dependency models which are easier to comprehend for non-technical experts. 

Finally, SecMoF provides an interactive method to view and analyse the models. It provides additional depth to the models compared to a static diagram or textual representations of attack trees. Using the interactive elements of the tool, an analyst may activate or deactivate a step within an attack model and view the impact it may have on a connected Dependency Model. Thereby assisting an analyst with establishing the severity of the impact an attack has on the system as a whole or its components. 

The main contributions of this paper are as follows:
\begin{enumerate}
  \item A novel way of representing Attack Trees using OIIRP format and SecMoF.
  \item A new tool \cite{maynard_2023} to convert SAND Attack Trees \cite{jhawar_2015} into FRIPP compatible models.
  \item Informative case studies of nine significant ICS intrusions \cite{maynard_2020}, translated from SAND Attack Trees into SecMoF. Two examples of malware attacks on CNI – the BlackEnergy malware and a cyber-attack on Ukraine's national power grid in 2015 – are discussed in detail. All nine models are made available to the readers in the SecMoF repository\cite{secmof}.
\end{enumerate}

The remainder of the paper is organised as follows. In Section~\ref{related-work} and \ref{background} we detail the Related Work and Background respectively. An overview of the proposed Intrusion Model Conversion Methodology follows in Section~\ref{model-convert}. A detailed analysis of two examples of malware attacks on CNI is offered in Section~\ref{s:example}. Finally, we discuss lessons learned in Section~\ref{s:discussion} and share our concluding remarks and plans for future work in Section~\ref{s:conclusion}.

\section{Related Work}\label{related-work}


In this section, we review the relevant publications from three areas: cyber-attack modelling, incident response modelling and risk assessment. We specifically explore the methods that are applied in the CNI context. We describe our chosen methods Attack Trees with Sequential AND (SAND), and Operations-Informed Incident Response Playbooks (OIIRP) in depth in the background section (\S\ref{background}). 

First, we look at the publications that focus on attack modelling in the CNI sector. 
Stergiopoulos \cite{stergiopoulos_2020} surveyed cyber-attacks on the oil and gas sector between 1990 and 2020, they analysed the attacks based on real-work reports and published demo attacks. They aligned their findings with the MITRE ATT\&CK knowledge base and identified common and subliminal attack paths that have been deployed against oil and gas.
Kumar \cite{kumar_2022} proposes using SAND Attack Trees modified to include graphics alongside the textual descriptions of the nodes to provide an easier experience for practitioners. They model three popular intrusions using this approach: Stuxnet, BlackEnergy, and Triton. Ebrahimi \cite{ebrahimi_2022} proposes a novel threat modelling methodology to construct attack paths which closely align with the asset. This work is focused on the automotive industry, specifically connected vehicles. 
Their use of Attack Trees is referred to as "Anti-Pattern Trees", which is used to derive threat rules which are aligned with their Data Flow Diagram that represents the system being analysed. 
Tan \cite{tan_2022} uses SAND Attack Trees to model two mock intrusions into a water reservoir testbed. The first scenario centres around a physical and network-based attack on the PLCs and HMI. The second describes a supply chain attack in which a 4G router has been compromised and is used as a foothold in the OT network. 
Kriaa \cite{kriaa_2012} proposes the use of Boolean Logic Driven Markov Processes (BDMP) to model the Stuxnet attack, which is a graphical modelling formalisation designed for safety and reliability assessments. A unique feature of Kriaa's approach is the dynamic modelling using a special relation called "triggers". This allows the capture of "Timed Security Events", which are events necessary for the success of the attack but are not under direct control of the attacker.

Second, we look at the commonly used IR modelling approaches. The incident response domain covers IR processes and Courses of Action (CoAs) that constitute countermeasures to cyber-attacks. Two popular IR formats are IACD and CACAO. 
Integrated Adaptive Cyber Defense (IACD) \cite{9557787} Playbooks were developed in 2014 and are maintained by the United States' Department for Homeland Security (DHS), among others. IACD playbooks are designed to provide a structured response, in combination with two other levels, workflow and local instances. Workflows are the machine-understandable codification of playbooks that enable automation of the steps, while local instances are the execution of tailored actions for a specific system. IACD use the XML format to store the Playbooks and BPMN for workflows.  

Created in 2017, Collaborative Automated Course of Action Operations (CACAO) for Cyber Security \cite{9557787} is maintained by OASIS. It uses the JSON format to represent playbooks. These playbooks aim to provide a precise structured definition of countermeasures, which can be automated and supported by a range of technologies and inter-organisational operations.
As highlighted by Shaked \cite{shaked_model_2022}, IR playbooks like CACAO and IACD often employ free-form high-level diagrams and tabular natural language descriptions which do not allow for a precise and structured representation of playbooks for a consistent and compatible modelling such as offered by OIIRP \cite{shaked_operations-informed_2023}.

Third, we briefly examine the existing methods for risk assessment in CNI. More detailed reviews of risk assessment methods are available in \cite{cherdantseva_2016} and \cite{chehri2021security}. Due to the rapidly changing threat landscape and the existence of evolving sophisticated cyber-attacks Kure \cite{kure_2022} have proposed a novel integrated cyber security risk management (i-CSRM) framework. That systematically identifies critical assets by using a decision support mechanism built on fuzzy set theory, and by predicting risk types through machine learning models. They evaluate their approach against a real case study of a critical infrastructure, in which they were able to correctly classify a denial-of-service attack, cyber espionage, and crimeware. 
Baggott \cite{baggott_2020a} presents a framework for risk analysis of cyber security and critical infrastructure procreation of the electrical power grid. They aim to link Hierarchical Holographic Model (HHM), Risk Filtering, Ranking and Management (RFEM) and Failure Modes Effects and Criticality Analysis (FEMCA) as a structure to qualitatively and quantitatively address the risk associated with the operation of the power grid. They identify several areas for further discussion and examination regarding the current electrical grid vulnerabilities, and to better understand the effects of a coordinated attack on the American power grid. Amro \cite{amro_2023} have proposed a novel risk assessment method for autonomous passenger ships, based on FEMCA (IEC 60812), and parts of the MITRE ATT\&CK framework. They then perform a risk assessment of the communication architecture based on the requirements of an autonomous ship. Allowing them to identify a group of metrics that estimate the impact of the risks and a set of countermeasures. Bolbot \cite{bolbot2020novel} also analyses the cyber risks for marine systems by expanding upon the Cyber Preliminary Hazard Analysis (CPHA). Like Baggott, Bolbot based their analysis on the safety focus methods and expanded them to include cyber risks.

Shaked \cite{shaked2023model} proposed TRADES (Threat and Risk Assessment for Design of Engineered Systems) a model-based methodology to support the design and assessment of systems' security aspects. TRADES fits within the System Security Engineering discipline, as a modelling tool to assist in mitigating risks during the design of systems. TRADES provides mechanisms to integrate, communicate and manage information regarding the system design concerning security risks throughout the development lifecycle. Critically, TRADES does this in a way which can be understood and shared between all skill levels from the board of directors, and security analysts, to systems engineers. 

Zografopoulos \cite{zografopoulos_2021} developed a novel threat modelling method and risk assessment that is designed specifically for cyber-physical energy systems. Their approach allows them to accurately represent the cyber-physical system elements, their interdependencies, as well as the possible attack entry points. What stands out from this work, is their unified approach of threat modelling and risk assessments, which complement each other.
Cyber risk assessments for electrical substations are used within the insurance industry, as described by Yang \cite{yang_2020}, which proposes a framework of premium calculation for cyber insurance businesses by modelling potential cyber intrusions and its expected mean time to restore power (MTTRP). 

Neither of the publications examined above attempts to bring together cyber-attack modelling with incident response modelling. The analysis of the related literature confirms that threat and response modelling are conducted separately in an uncoordinated manner, failing to learn and benefit mutually from each other. 

Of the examined related work, we identified three which are discussed below as most closely related. First, is Shaked \cite{shaked2023model} TRADES methodology, in which they can capture the cyber risk information of a system throughout its lifecycle and present it in an easy-to-understand fashion without losing the technical provenance of the risk. Second, Yang \cite{yang_2020} models cyber intrusions and their impact on the critical national infrastructure in terms that may be broadly understood as "mean time to restore power". Third, Zografopoulos \cite{zografopoulos_2021} proposed a threat modelling framework for CNI, in which they defined a set of unique CNI elements. Zografopoulos' framework allows mapping CNI elements interdependencies to create a high-fidelity model.
Both Yang \cite{yang_2020} and Zografopoulos \cite{zografopoulos_2021} cannot show the impact a threat might have on the system as a whole while maintaining the risk provenance when in discussions with non-technical stakeholders. While Shaked \cite{shaked2023model} may maintain provenance, they don't attempt to map the impact of the risk. Moreover, neither of them is compatible with both a risk assessment model and a threat model, thereby losing some of the potential lessons learned by performing each task independently.

There are recent attempts to provide unified models or representations that combine the three perspectives, namely, cyber-attack, incident response and risk assessment. Xiong et al. \cite{xiong2022cyber} propose a domain specific language for assessing the cyber security of enterprise systems by simulating attacks with respect to the system architecture. This language does not account for aspects of incident response. Mouratidis et al. \cite{mouratidis2023modelling} offer a hybrid representation of risks, attacks and incident response without proper process orientation and with fragmented, partial views on the affected system. None of the related work allows simultaneous interplay between attack and incident response based on a common model of the affected system. A recent systematic literature review of threat modelling in industrial control systems context establishes a gap in communicating threats between pertinent stakeholders \cite{khalil_2024}, to which the disparate modelling techniques contribute.

\section{Background}\label{background}

In this section, we provide an introduction to the two modelling methods integrated, analysed and compared in this work: Attack Trees with Sequential AND (SAND) and Operations-Informed Incident Response Playbooks (OIIRP).

\subsection{Attack Trees with Sequential AND}\label{sand-attack-trees}

Threat modelling allows system designers to identify potential countermeasures by understanding how an attacker may exploit a weakness within a system. To help identify the countermeasures, one may put together a list of steps needed to break a system. Attack trees are designed to allow for identifying different ways in which a system or process can be attacked systematically. Attack trees \cite{schneier_1999} traditionally start with an attacker's goal at the top, called a root node, while subsequent child nodes represent the attacker's sub-goals. The nodes are connected using disjunctive (OR) and conjunctive (AND), with the leaves of the tree representing the attacker's actions. 

Attack Trees with Sequential AND (SAND) is an enhancement of attack trees defined in 2015 that has been formally described in \cite{jhawar_2015}. SAND adds another relationship operator along with the original (OR, AND) called the "sequential conjunctive (SAND)". This ensures compatibility with future and past works based on this formalisation. The SAND operator has been in use long before the formal definition. 

Attack Trees are a popular formalisation for use within the critical infrastructure industry due to the similarity of attack trees to fault trees \cite{nuclear_regulatory_commission_fault_1981}, which are used in industrial process design to identify faults. Attack trees are often represented as tab-indented text, making them easy to comprehend both the raw and visual forms.

\begin{figure*}[t]
  \begin{minipage}[l]{0.33\textwidth}
\begin{lstlisting}[basicstyle=\tiny,tabsize=1,frame=single]
Become Root:OR
	No-auth:SAND
		Gain user privileges:SAND
			ftp
			rsh
		Local Buffer Overflow
	Auth:AND
		ssh
		rsa
\end{lstlisting}
  \end{minipage}
  \begin{minipage}[r]{.66\textwidth}
    \includegraphics[width=\textwidth]{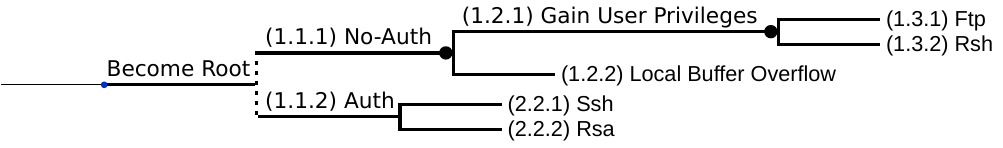}
  \end{minipage}
  \caption{SAND Attack Tree example (Left: Tab intended plain text. Right: Visualised) \cite{maynard_2020}}
  \label{fig:sand-example}
\end{figure*}

An example attack tree based on the formalisation from \cite{jhawar_2015} using SAND is shown in Figure~\ref{fig:sand-example}, the textual tab-indented tree is shown on the left, and a graphical tree is shown on the right. The example shows the goal of the attacker is to "Become Root", there are two ways for them to do this, without authentication \underline{or} with authentication:
\begin{enumerate}
  \item \textbf{No-Auth (1.1.1)}: Unauthorised access by \underline{first}:
  \begin{enumerate}
    \item Gaining user privileges (1.2.1) by \underline{first} exploiting the FTP service (1.3.1) \underline{then} exploiting the restricted shell, rsh (1.3.2). 
    \item \underline{Then} perform a local buffer overflow (1.2.2) to become root.
  \end{enumerate}
  \item \textbf{Auth (1.1.2)}: Authorised access by using SSH (2.2.1) \underline{and} having access to the private RSA key (2.2.2) associated with the user's public key on the server.
\end{enumerate}

In this example, the SAND operator is used twice. First "No-auth (1.1.1)" specifies that an attacker will first need to "Gain user privileges (1.2.1)" on the server, and then perform a "local buffer overflow (1.2.2)" which will give them root access. 

The second time the SAND operator is used is on the "Gain user privileges (1.2.1)" branch. This means that first, the attacker must exploit a vulnerability in the "FTP (1.3.1)" service to give them access to the restricted shell, "RSH (1.3.2)", which they will also need to exploit to "Gain user privileges (1.2.1)" on the server. 
\subsection{Operations-Informed Incident Response Playbooks}\label{oiirp}

An Operations-Informed Incident Response Playbook (OIIRP)~\cite{shaked_operations-informed_2023} combines a Formalised Response to Incident Process Playbook (FRIPP) \cite{shaked_model_2022} and related Dependency Models \cite{slater_2016} into a unified approach for capturing both IR and the operational impact in one modelling environment. In OIIRP, an IR playbook may be associated with its operational context and its impact on the operational context defined within a dependency model. We explain all relevant concepts below. 

\subsubsection{Formalised Response to Incident Process Playbook}\label{process-model}

\begin{figure}[t]
    \includegraphics[width=\columnwidth]{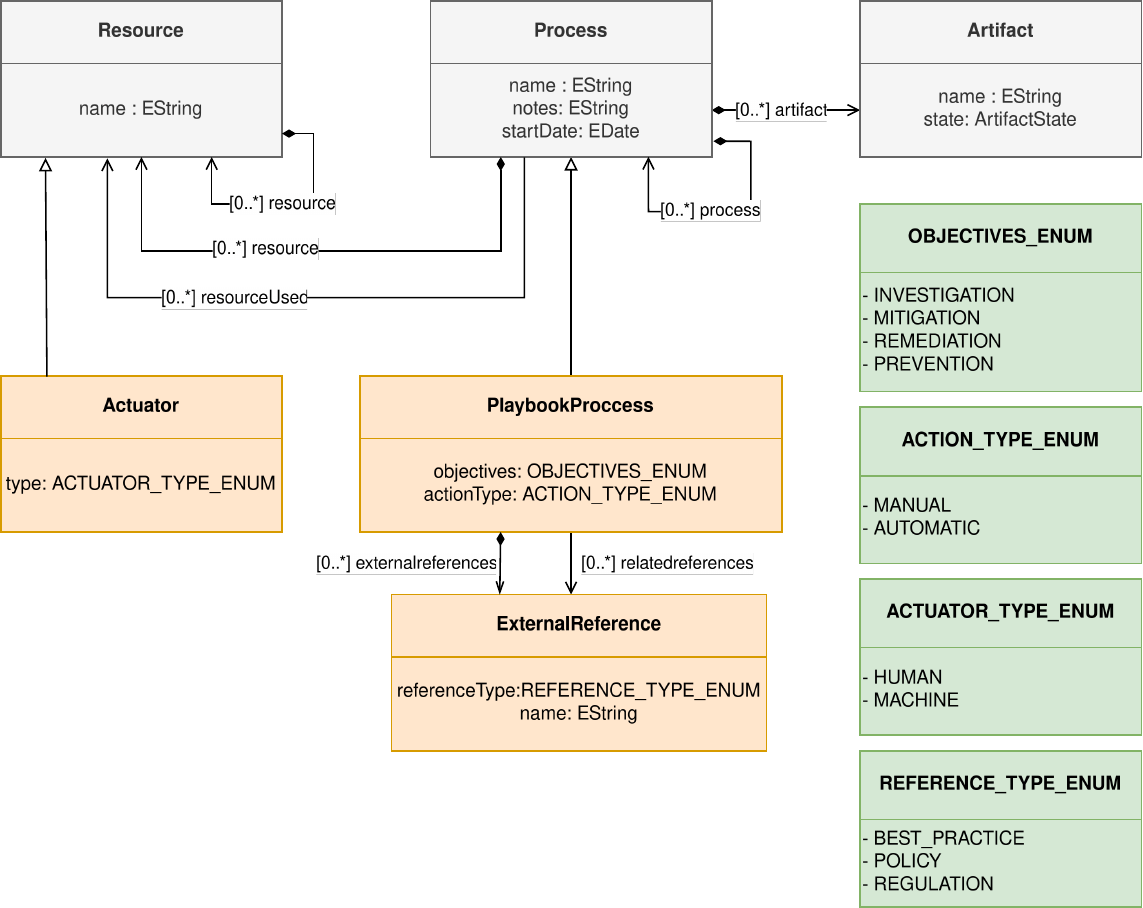}
    \caption{The Meta-Model for Formalised Response to Incident Process Playbook (FRIPP).}
    \label{fig:fripp-meta}
\end{figure}

Shaked \cite{shaked_model_2022} proposed and defined a formal, model-based approach for cyber security incident response playbooks called Formalised Response to Incident Process Playbook (FRIPP), based on the PROVE Tool \cite{shaked_prove_2022} created for designing and analysing process descriptions. The FRIPP meta-model is shown in Figure~\ref{fig:fripp-meta}. FRIPP defines a formal representation of IR playbooks as follows: 

\begin{enumerate}
  \item Playbook Process: The main element describing an IR process in any hierarchy, which contains a set of interconnected Processes representing actions.
  \item Actuator: Identifying a person or a machine responsible for executing a Process.
  \item External Reference: Relates a Playbook to external references for guidance or demonstrating compliance with standards.
\end{enumerate}

\begin{figure*}[t]
    \includegraphics[width=\textwidth]{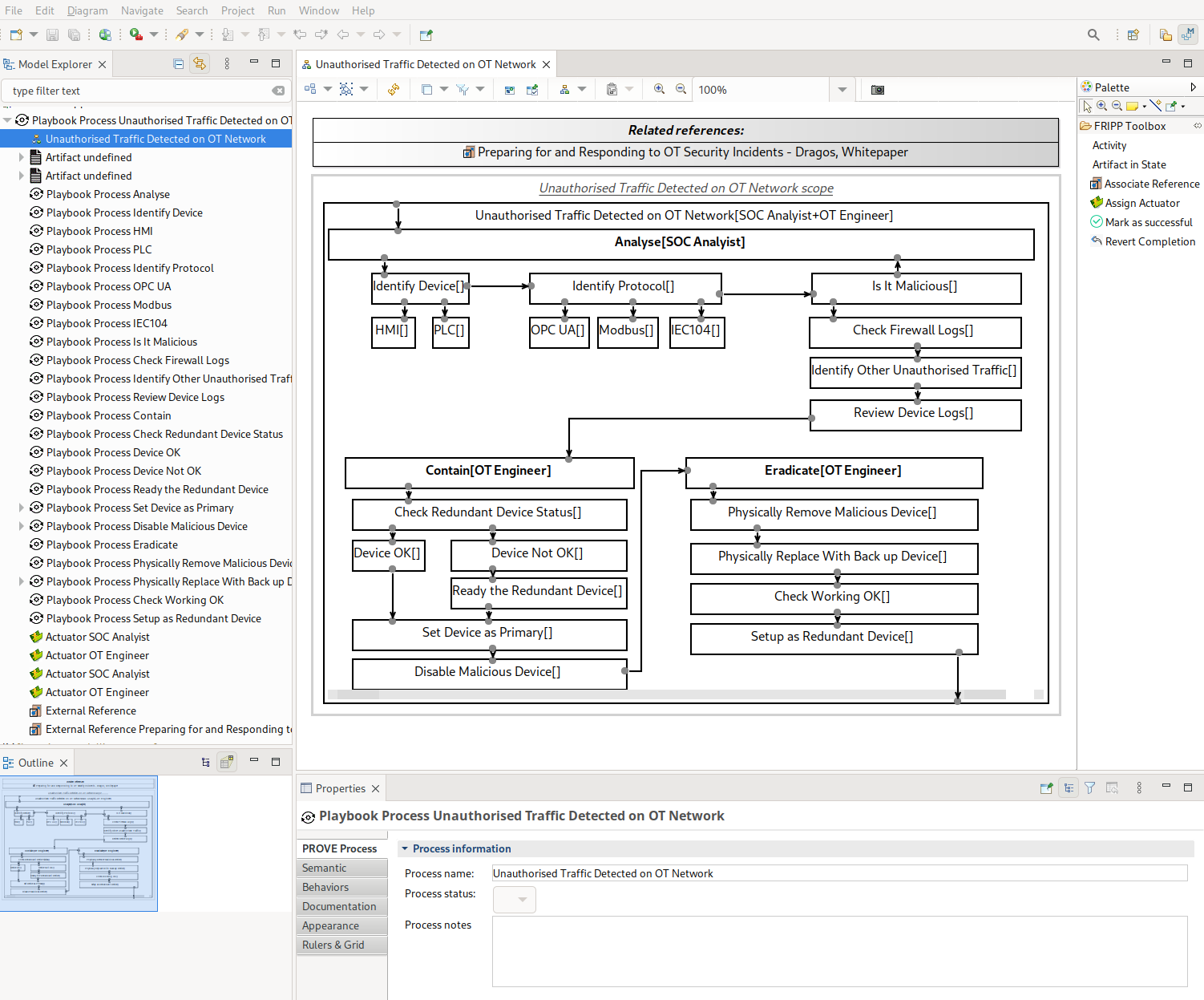}
    \caption{Unauthorised Network Traffic on OT Network FRIPP playbook designed using SecMoF.}
    \label{fig:fripp-example}
\end{figure*}

The benefit of FRIPP over other modelling methods is the concise expressiveness, which engages practitioners who do not have a background in modelling. An example of an IR playbook designed in CACAO and converted into the FRIPP model within the SecMoF is presented in \cite{shaked_model_2022}. A related IR playbook is shown in Figure~\ref{fig:fripp-example} - this is an example of a FRIPP model for unauthorised network traffic detected on an OT network. This playbook is specific for an example CNI and follows a typical ordering of steps: Analyse, Contain, and Eradicate. Under each of the main steps, there are several others, some only have one path, and some are dependent on the state of the system. For example, under the Analyse step, there is a step titled "Is it  Malicious". As this playbook may be triggered by an automated IDS, the SOC analyst would have to check the traffic alerted was not a false positive, and if it is then the playbook would end at the analysis stage. Figure~\ref{fig:fripp-example} demonstrates the SecMOF modelling environment and its underlying model-driven approach: the Model Explorer tab on the left allows to navigate the model and locate specific element; the diagrammatic representation at the centre is generated using queries with respect to the model; and the Properties tab at the bottom shows that each element has a set of properties associated with it.

\subsubsection{Dependency Modelling}\label{dependency-modelling}

Dependency Modelling (DM) \cite{slater_2016} is a standard published by the Open Group in 2016 for constructing quantitative data models for risk management of organisational inter-dependencies. Dependency models incorporate physical, cyber, geographical, and logical dependencies, in a positive top-down model. By focusing on what is required for a system to operate successfully, as opposed to what faults may occur, ensure effort is directed towards areas which are most critical to continued operation. 

\begin{figure}[t]
\includegraphics[width=\columnwidth]{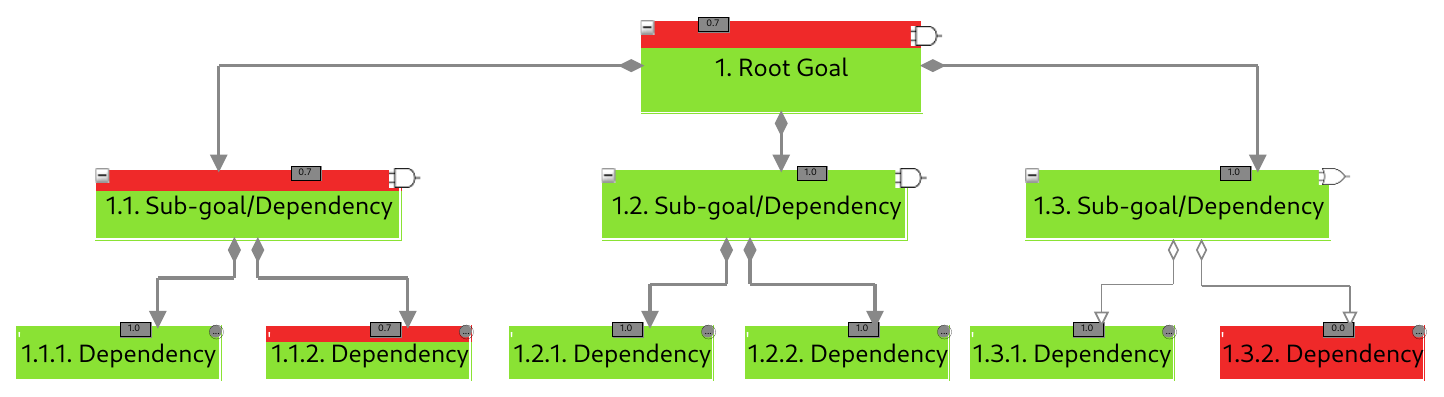}
\caption{An Example of a Dependency Modelling Artefact.}
\label{fig:dm-example}
\end{figure}

A dependency model is represented as a directed graph in a tree-like structure, with each node representing a dependency or goal. In DM, each node is known as a Paragon, and edges as dependency relationships. Table~\ref{tab:Paragon} describes the attributes of a Paragon. An example of a dependency model is shown in Figure~\ref{fig:dm-example}. The top of the graph is the "Root Goal", and each of its children nodes makes up a "sub-goal" or "dependency" required for the success of the root goal. A relationship may be \textbf{OR-typed} - represented in Figure~\ref{fig:dm-example} by a hollow diamond at the dependant Paragon end and empty arrowhead at the sub-goal end (see Paragon 1.3), or \textbf{AND-typed} – represented in Figure~\ref{fig:dm-example} by a filled diamond and arrowhead (see Paragon 1.2). 

An AND relationship requires several nodes to function properly without breaking, while the OR relationship requires only that either of the nodes functions. The AND relationship will fail if any node fails (increases the risk to the parent), whereas the OR relationship will succeed if at least one child succeeds (reduces the risk to the parent).

In the example Figure~\ref{fig:dm-example} many of the Paragons have a state '1' (green), meaning they are Ok, while node 1.3.2 is marked as failed with a state of '0' (red) and node 1.1.2 has a state of 0.7. Due to their relationships to their parent nodes, both states affect the root node differently. 
Because node 1.3 has an OR-typed relationship with its children, its state is still '1' because node 1.3.1 has a state of '1'.
While node 1.1.2 has a state of '0.7' and its parent relationship is AND-typed, this state is propagated upwards to the parent, then up to node 1. as this is also AND-typed. 

\begin{table*}[t]
\tiny{}
\caption{Attributes of a Dependency Model Paragon \cite{slater_2016}.}
\begin{tabular}{p{1.7cm}p{5cm}p{5cm}}
\toprule
\textbf{Feature} & \textbf{Description} & \textbf{Example} \\
\midrule
Name or \newline Label & The name represents the positive "success" entity we want to associate with the Paragon. It should not hide its importance. If you mean nobody is to die, don't say "safety addressed", which could mean anything, say "nobody dies" & "trip successful, car OK".\\
\hline Positivity & A Paragon must only state the goal, not problems. For linguistic reasons, some goals can only be expressed as the absence of something. & It is okay to have a goal that means no earthquakes, or no financial downturns. \\ 
\hline Abstraction & A Paragon is abstract, not material. So a goal might be to "Keep out the bad guys", not "an access control system". Because we can have an access control system yet still fail to keep bad guys out. & "Keep out the bad guys" \\
\hline No \newline Box-Ticking & We want to identify risks, so if a Paragon meant the existence of an access control system, this would give it a 100\% probability of success, because it exists. & N/A \\
\hline Strong \newline Definition & Paragons have a meaning expressed through their definition. Don't confuse the name with the definition, the name is a shorthand reminder. The definition needs to be precisely defined. & To prevent attackers or their agents from entering company premises, or introducing malicious software into company computers or networks at any time between midnight January 1st 2011 to midnight January 1st 2012, or to detect their presence in a sufficiently timely manner as to be able o prevent any of the losses cataloged. \\
\hline Consistency & Each Paragon is defined by its relationship with its dependencies. & Either 'AND' or 'OR' relationship. \\ 
\hline Aspect & There is not a one-to-one correspondence between a Paragon and any physical entity. If we are describing a nuclear power station, there are several aspects. & One would not be "the nuclear power station": 
\begin{itemize}
  \item success as a source of electricity 
  \item prevents the release of radioactive particles into the environment 
  \item its effectiveness as a supplier of local employment 
  \item the speed with which it could respond to sudden demands 
  \item its cost-effectiveness as an energy source 
  \item its resilience to nation-state cyber-attacks 
\end{itemize}\nointerlineskip \\
\hline States & The degree of availability. A Paragon has at least two states. The states shall always be arranged in order from worst to best. Sometimes verbs are used as shorthand to describe which state a Paragon is in. & Failure state 0, and complete success as state 1. With anything between 0.1 to 0.9 as a percentage of success.\\
\bottomrule
\end{tabular}
\label{tab:Paragon}
\end{table*}

\subsubsection{Linking Playbooks with Dependency Models}\label{ss:back:oiirp:linkdm}

\begin{figure*}[t]
    \includegraphics[width=\textwidth]{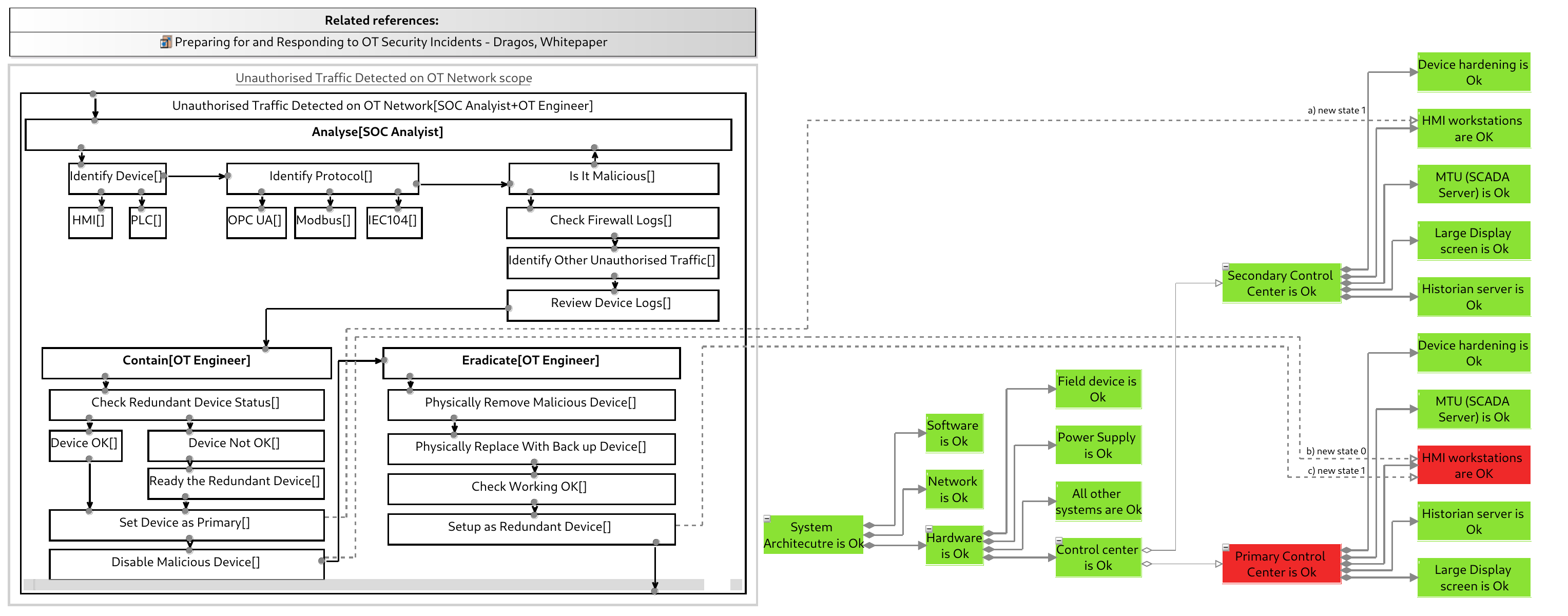}
    \caption{Linking the FRIPP Playbook "Unauthorised Traffic Detected on OT Network" with a system dependency model in SecMoF.}
    \label{fig:unatht-dm}
\end{figure*}

The unique feature of OIIRP is the linking of FRIPP playbook steps to specific DM Paragons and the change of their state. It is possible to link any number of Paragons to any number of FRIPP steps. Figure~\ref{fig:unatht-dm} shows the same FRIPP playbook as described in Section~\ref{process-model} linked to a sample SCADA dependency model. The dependency model describes a simple CNI system with two control centres along with their sub-dependencies.

In this example, there are three links from the FRIPP playbook to the dependency model. Table~\ref{tab:linking} details which step of the playbook is linked to which Paragon, and the resulting state change. Step a) ensures that the secondary control centre's HMI is operational and then sets the device as primary to ensure business continuity. Step b) disables the primary control centre's HMI, in which unauthorised traffic was detected. Finally c) after the device has been replaced with a new trustworthy one, the HMI is set up as a redundant device. This example shows how IR steps have an impact on the states of the dependency model. 

\begin{table}[t]
\caption{Summary of impact the FRIPP playbook has on the system dependency model.}
\label{tab:linking}
\begin{tabular}{p{4cm}p{4cm}p{1.2cm}}
\toprule
\textbf{FRIPP Step} & \textbf{DM Paragon Impact} & \textbf{State} \\
\midrule
a) Contain/Set Device as Primary & Secondary Control Center is Ok/HMI Workstations are OK & $1  \longmapsto 1$ \\
b) Contain/Disable Malicious Device & Primary Control Center is Ok/HMI Workstations are OK & $1 \longmapsto 0$ \\
c) Eradicate/Setup as Redundant Device & Primary Control Center is Ok/HMI Workstations are OK & $0 \longmapsto 1$ \\
\bottomrule
\end{tabular}
\end{table}

\section{Intrusion Model Conversion Methodology}\label{model-convert}

The purpose of the Security Model Converter (SMC) tool that we created in this project is to translate free-form SAND attack trees into FRIPP models, which could be viewed and modified using SecMoF. Once loaded into SecMoF, the attack tree FRIPP models are manually modified to include actuators for each stage of the intrusion and any associated references. 

After this point onwards, converted attack trees differ from traditional FRIPP models (as discussed in Section~\ref{meth:meth}) and are referred to in the rest of the paper as Compatible Intrusion Models (CIM). As CIMs are based on the FRIPP meta-model (see Figure\ref{fig:fripp-meta}) they only differ in the way they are being used. Having attack trees represented as CIMs allows mapping intrusion steps to a Dependency Model to identify the impact an attack and separate steps have on a system (or its components) as demonstrated in Section~\ref{s:dependency-model-analysis}. 

Section~\ref{meth:smc} explains how the SMC tool works, while Section~\ref{meth:meth} details the conversion steps from a SAND Attack Trees to a CIM and the differences between FRIPP and CIM. 

\subsection{Security Model Converter}\label{meth:smc}

The SMC\cite{maynard_2023} is an Open-Source command-line application written in GoLang that converts security model data to different formats. SMC supports conversion between SAND Attack Trees, Dependency Models, and FRIPP. 

Attack trees are typically written as tab-indented plain text (See Section~\ref{sand-attack-trees}) while more complex models such as FRIPP use XML to capture more detail. SMC uses GraphML as an intermediate format to allow conversion of the different formats, since GraphML is a widely used format it may be used to import/export the models into other tools, including and beyond SecMoF. 

Table~\ref{tab:smc-io} breaks down the supported formats and the types of models that SMC can convert between.
SMC works alongside SecMoF as an independent tool which is used to help manage the datasets used by SecMoF. SMC does not support verification of the models as this is done within SecMoF \cite{shaked_model_2022,shaked_operations-informed_2023}.    

\begin{table}[t]
\caption{Input and Output formats of the Security Model Converter.}
\begin{tabular}{p{1.3cm}p{0.2cm}p{0.2cm}p{0.2cm}p{0.2cm}p{0.2cm}p{0.2cm}p{0.2cm}p{0.2cm}}
\toprule
\multirow{2}*{\textbf{Format}} & \multicolumn{2}{c}{\textbf{Tab Indented}} & \multicolumn{2}{c}{\textbf{GraphML}} & \multicolumn{2}{c}{\textbf{SecMoF}} & \multicolumn{2}{c}{\textbf{iDepend}} \\
       & In    & Out  & In      & Out & In    & Out  & In      & Out \\
\midrule
SAND & Y&\textbf{N} & Y&Y & -&-
& -&- \\
DM & -&- & Y&Y & Y&Y &
Y&\textbf{N} \\
FRIPP & -&- & Y&Y & Y&Y &
-&- \\
\bottomrule
\multicolumn{9}{c}{\textbf{Y}: Yes;    \textbf{N}: No;    \textbf{-}: Not Applicable;}\\
\bottomrule
\end{tabular}
\label{tab:smc-io}
\end{table}

\subsection{Conversion Methodology}\label{meth:meth}
\begin{figure}[t]
    \includegraphics[width=\columnwidth]{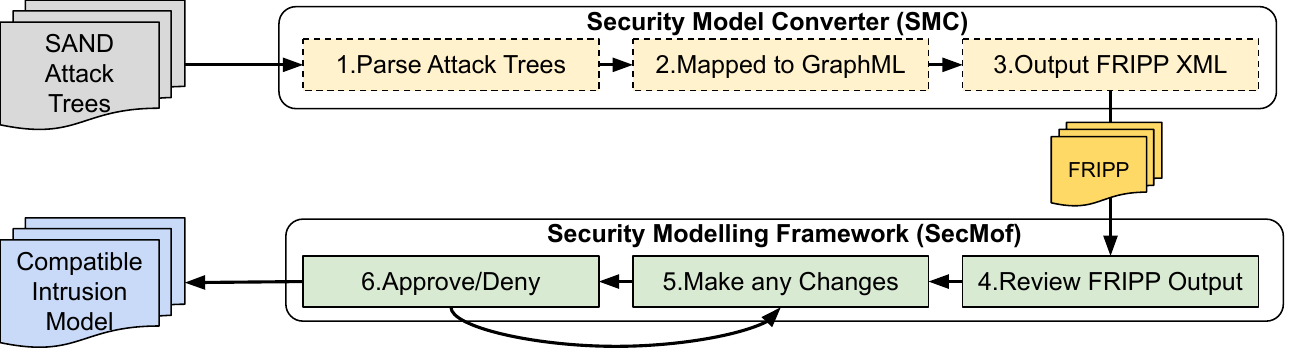}
    \caption{Model conversion methodology showing the two tools. The dashed boxes indicate an automated action, while the solid is a manual intervention.}
    \label{fig:intrusion-methodology}
\end{figure}

Figure~\ref{fig:intrusion-methodology} depicts and describes the steps that need to be taken to convert a SAND Attack Tree model into a CIM. The steps are broken down into two stages based on which tool they take place:
\begin{enumerate}
  \item \textbf{SMC}: Takes an initial tab indented attack tree and maps it into a GraphML representation. From this representation, an attack tree is generated in the form of a FRIPP playbook, which is formatted as XML and supported by the SecMoF tool. These steps are automated by the SMC tool.
    \begin{enumerate}
        \item Read the source model into memory, which is stored in an interchangeable GraphML format.
        \item Transform the GraphML model in memory into SecMoF compatible XML format as either FRIPP or Dependency Model and write the output to disk. 
    \end{enumerate}
  \item \textbf{SecMoF}: After conversion to FRIPP, the first step is to review the outputted files within SecMoF to verify they can be loaded correctly. Then the contents of the trees are manually reviewed. An analyst could take advantage of SecMoF's features to allow for structured analysis of a converted model. Typically, the following changes could be made:
  \begin{enumerate}
    \item Assign an actuator and set its value to Automatic, Manual, Dual, or Unknown.
    \item Add associated references.
    \item Map intrusion steps to an associated Dependency Model.
    \item Resize shapes and arrows in the diagram for easier comprehension.
    \item SecMoF auto colours steps depending on actuator types. 
  \end{enumerate}
  Finally, a CIM is accepted and approved, or denied and requires a revision. If denied, changes are then made iteratively until the model is of satisfactory detail and accuracy. 
\end{enumerate}

\subsubsection{Differences Between CIM and SAND Attack Tree}\label{meth:cim-sand}

\begin{figure}[t]
    \includegraphics[width=\columnwidth]{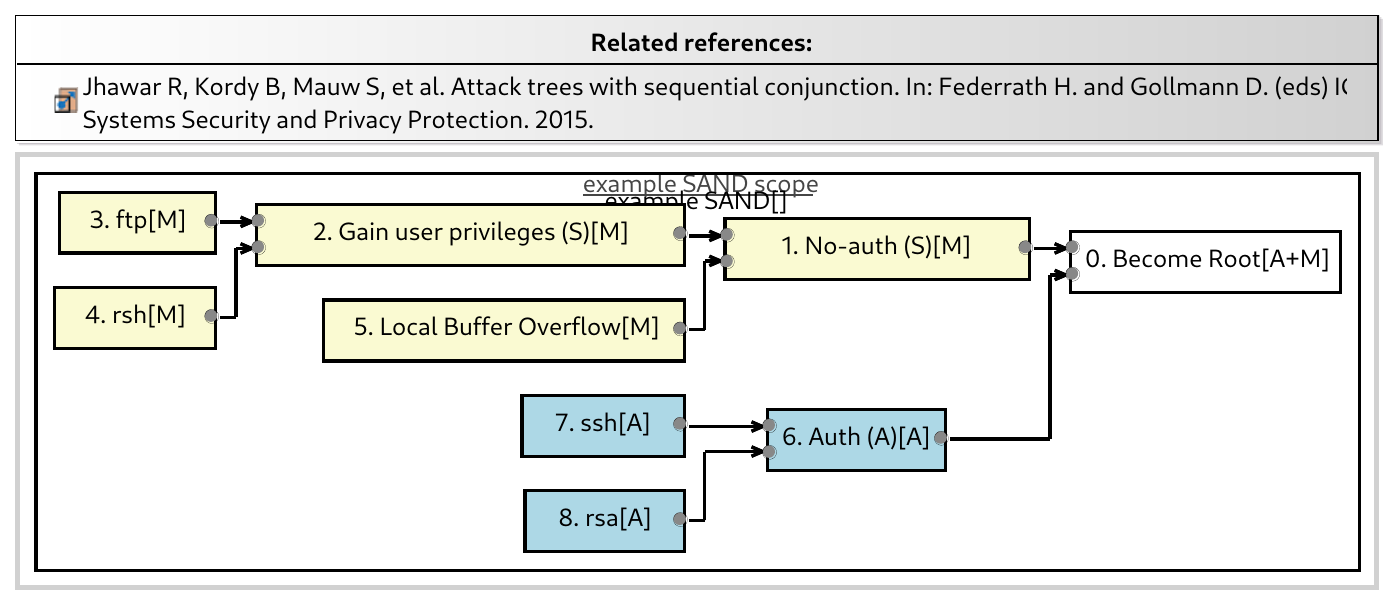}
    \caption{Example SAND Attack Tree represented as a Compatible Intrusion Model.}
    \label{fig:sand-cim}
\end{figure}

There are several differences between SAND Attack Trees and CIM in terms of the representation and the data they contain. Figure~\ref{fig:sand-cim} shows an example CIM which is a conversion of the SAND Attack Tree from Figure~\ref{fig:sand-example} discussed in Section~\ref{sand-attack-trees}. Additional information is depicted in the diagram, including Related references, using the 'Related references' container; and the types of actions (e.g. manual or automatic), using '[A]' for automatic, '[M]' for manual, or '[A+M]' for both.

Table~\ref{tab:cim-sand} shows the detailed comparison of the information contained in a SAND attack tree and a CIM.

\begin{table}[t]
\caption{Features of a Compatible Intrusion Model Compared to SAND Attack Trees}
\begin{tabular}{p{3cm}p{4cm}p{4cm}}
\toprule
\textbf{Feature} & \textbf{CIM} & \textbf{SAND} \\
\midrule
Related References & Included in Model & External to Model \\
Actuators & Assigned to a specific node as Manual, Automatic, Dual, or Unknown. & N/A \\
Diagram Direction & Right-to-Left & Left-to-Right \\
Relationship Operators & AND, OR & SAND, AND, OR \\
Node Numbering & Yes & Unofficially \\ 
\bottomrule
\end{tabular}
\label{tab:cim-sand}
\end{table}

In our experiments with model conversion, each of the SAND attack tree models that were converted into CIM required mapping to their respective 'Related references'. This is due to SAND not supporting this field, SAND is typically accompanied by an additional document which would contain references and further explanation of the model. For the models described in Section~\ref{s:example-intrusion-analysis}, the related references were taken from the original paper \cite{maynard_2020}. 

Actuators, as defined by FRIPP, identify a person or a machine responsible for executing a process. In the context of CIMs, each step of the intrusion may have one or more actuators. Based on our analysis of the SAND models, we have determined four actuators:
\begin{itemize}
   \item \textbf{Manual}: Requires input from a human, either the attacker or victim.  
   \item \textbf{Automatic}: An action can be performed by a script or program, either adversary or benign.
   \item \textbf{Dual}: Both manual and automatic actions.
   \item \textbf{Unknown}: The action type is unknown or/and no reliable source of information may be found.
 \end{itemize}

SAND Attack Trees place an attack objective at the root node with child nodes shown underneath them. The original SAND intrusions representations in \cite{maynard_2020} display a model left-to-right, with the root node on the left. This differs from traditional SAND Attack Trees and was done this way due to page restrictions and limitations of the graph visualisation library used to create the diagrams. The diagram direction of a CIM is opposite to the original SAND models with the root goal being on the right and children nodes on the left. When working on the models it became easier to comprehend the steps when displayed this way. This is because an intrusion is typically composed of small actions chained together that result in the successful exploitation of the victim. Chaining steps like this align with the ICS Cyber Kill Chain \cite{assante_2015}.

In our case with CIMs we change the representation direction. By placing the lowest-level actions on the left, we identify the 'low-hanging fruit' in terms of countermeasures that could be overcome by an attack, as these nodes on their own cannot result in successful exploitation of an intrusion, but contribute to it. One may consider the furthest left nodes as the easiest actions for an attacker to perform and the furthest right as the hardest, as this requires chaining multiple attack actions together.

The CIM supports Sequential AND by numbering the nodes combined with the AND operator, this is indicated by a \texttt{(S)} at the end of the node description. The reader may determine the ordering of the steps by following the numbering system. 
Number of the nodes takes inspiration from Integration Definition's \cite{IDEF_1999} family of modelling languages, specifically from IDEF0 which is a method for describing manufacturing functions. Each of the functions represented included a number to indicate the ordering of the steps. This numbering aligns well with ordering attack steps. SAND does not enforce the numbering of nodes, however, many representations of attack trees will number the nodes to help the reader comprehend the attack paths. 

\section{Examples of Converted Models for High Profile Attacks on CNI}\label{s:example}
The detailed analysis of nine significant cyber-attacks on CNI modelled using SAND methods is presented in \cite{maynard_2020}. The attacks examined include Stuxnet, Havex, BlackEnergy, German Steel Mill, Duqu2.0, Ukraine Power Outage 2015 and 2017, CrashOverride, and TRISIS.  We converted all nine attacks from Maynard \cite{maynard_2020} to CIM and these CIM models are available in the SecMoF project repository\cite{secmof}.  

In this paper, due to the limited space, we discuss in detail only two high-profile attacks out of the nine re-modelled attacks - the 2015 Ukraine power outage and the BlackEnergy malware. The models of the other seven attacks are presented in Appendix~\ref{appendix:remain}. We examine the advantages and new insights offered by our proposed modelling approach providing specific examples based on the BlackEnergy malware and Ukrainian power distribution systems attacks, and then summarised findings from the other seven models in Section~\ref{overview-insight}.

The remainder of this section is broken down into two categories: 
\begin{enumerate}
  \item \textbf{Intrusion Analysis (\S\ref{s:example-intrusion-analysis})} of the BlackEnergy Malware attack and the 2015 Ukraine Power Outage attack.
  \item \textbf{Dependency Analysis (\S\ref{s:dependency-model-analysis})} where we discuss system dependencies in the affected infrastructure and then map the BlackEnergy (\S\ref{BlackEnergy_dm}) and the Ukraine outage (\S\ref{ukraine_dm}) to the SCADA dependency model.
\end{enumerate}

\subsection{Intrusion Analysis}\label{s:example-intrusion-analysis}

The terminology used in this paper to describe the intrusion steps follows the PrEP framework proposed by Herr \cite{herr_2013}. PrEP classifies malware and cyber weapons based on the different pieces of malicious code that constitute them. The framework consists of three components: 

\begin{enumerate}
  \item \textit{A Propagation Method (\underline{Pr})}: Is how the malware propagates itself between machines.
  \item \textit{An Exploit (\underline{E})}: Is code designed with a malicious purpose, to compromise some aspect of a software system, that allows third parties to cause unintended operations of consequences.
  \item \textit{A Payload (\underline{P})}: Is code with a malicious purpose whose delivery and execution are the goals of any piece of malware.
\end{enumerate}

The three components are designed to be modular and all malware is perceived as comprising of these three fundamental components. PrEP attempts to overcome the issues with the current vague and ambiguous terminology such as worm, trojan, and virus and focuses on the actual characteristics of malware. Herr \cite{herr_2013} used the framework to analyse two major examples of malware, Stuxnet and Red October. 

When referring to a specific step within a CIM model, they are presented within quotes, no italics. e.g. "1. Propagate" When referring to a DM Paragon they will be presented in verbatim, no quotes. e.g. \texttt{Alerting Ok}.

\subsubsection{BlackEnergy Malware Attack}\label{BlackEnergy}
\begin{figure*}[t]
    \includegraphics[width=\textwidth]{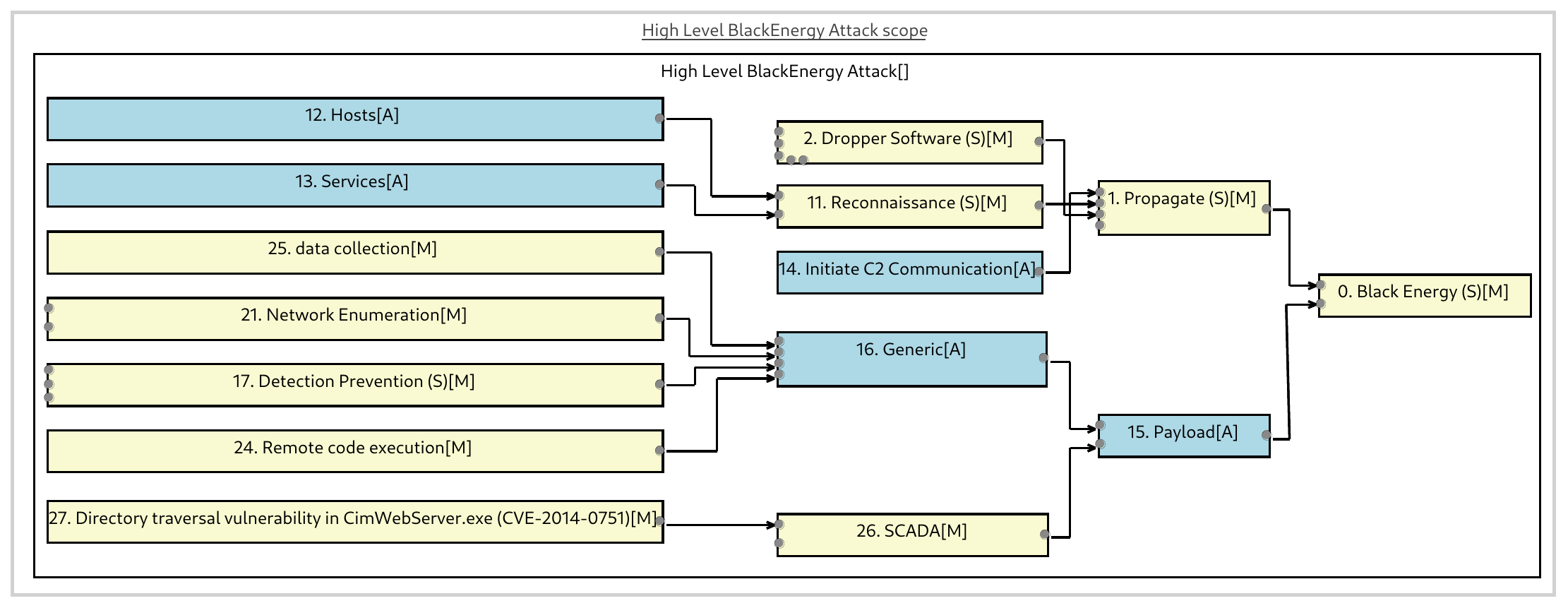}
    \caption{High Level BlackEnergy Attack Modelled in CIM.}
    \label{fig:be-hl-cim}
\end{figure*}

BlackEnergy is the malware that has been associated with several intrusions dating back to 2007\cite{ics-cert_2014,constantin_2014,geiger_2020}. The malware has been continually used due to its flexible architecture allowing the use of plug-ins. BlackEnergy has often been used for information harvesting and espionage. In this paper, we focus on BlackEnergy-based attacks on Human Machine Interface (HMI) in CNI, and, particularly, its role in the 2015 Ukrainian power outage. Figure~\ref{fig:be-hl-cim} shows the high level overview of BlackEnergy modelled using CIM, with Listing~\ref{fig:be-hl-text} showing the tab indented equivalent. 

At a high level, the attack can be broken down into two stages. The first stage is propagation, in which the adversaries manually deploy dropper software on the victim's network resulting in the compromise of the domain controller and ultimately gaining privileged access to the whole network which they use to perform reconnaissance and initiate command and control (C2) communications. The second stage is payload deployment, as BlackEnergy is often used for different scenarios, it could deploy generic payloads such as network enumeration or remote code execution. However, it had payload support for SCADA systems, specifically the ability to exploit a directory traversal vulnerability in the WebView component of a GE Intelligent Platforms Proficy HMI/SCADA system. By exploiting this the attacker can execute arbitrary code on the server and gain complete control over the system.

\begin{minipage}{\linewidth}
\begin{lstlisting}[basicstyle=\tiny,tabsize=1,frame=single,breaklines=true,caption=High level BlackEnergy attack in tab-indented format.,label=fig:be-hl-text]
Black Energy : SAND
	Propagate : SAND
		Dropper Software : SAND
		Reconnaissance : SAND
			Hosts : OR
			Services : OR
		Initiate C2 Communication
	Payload : OR
		Generic 
			Detection Prevention : SAND
			Network Enumeration : OR
			Remote code execution 
		SCADA 
			Directory traversal vulnerability in CimWebServer.exe (CVE-2014-0751)
\end{lstlisting}
\end{minipage}

\begin{figure*}[t]
    \includegraphics[width=\textwidth]{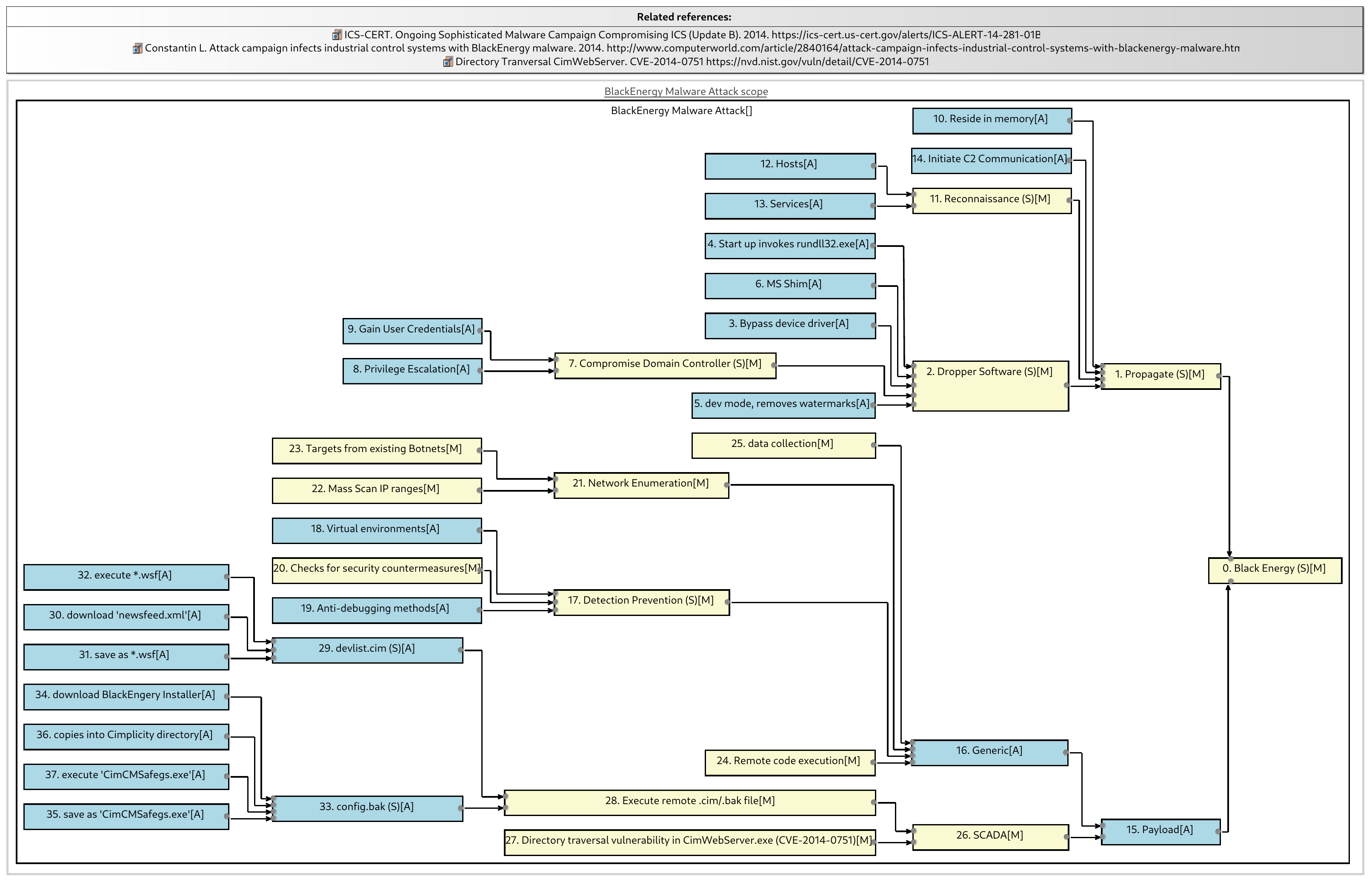}
    \caption{Full BlackEnergy Attack Modelled in CIM.}
    \label{fig:BlackEnergy-fripp}
\end{figure*}

Figure~\ref{fig:BlackEnergy-fripp} shows the full CIM representation of the BlackEnergy Malware attack. There are three primary sources of information for the industrial variation of BlackEnergy which are captured in the "Related references" section at the top of the diagram. SecMoF allows the association of related references within an attack model.  Related references are a novel feature of FRIPP, and was identified in \cite{shaked_model_2022} that other models such as CACAO, IACD, and SAND do not have, yet, in the context of incident response and threat modelling, related references become an important trait. 

There are 37 steps: 15 Manual steps (shaded in yellow) and 22 Automatic steps (shaded in blue). One of the advantages of CIM in contrast to SAND is the assignment of the action types that will be automatically colour-coded in SecMoF. The manual interactions are due to the original design of the malware, which targeted political espionage, meaning that a fully automated attack would be detected with a higher probability. Therefore, by design, the malware waits for input from the attackers before performing its automated actions. Once the manual interaction has taken place the remaining operations are automatically performed.

Compared to the analysis of the Ukraine intrusion in Section~\ref{ukraine2015}, BlackEnergy contains more technical automated steps, e.g. all steps from "28. Execute remote .cim/.bak file[M]" onwards. The children of node 28 details the automated steps taken to compromise the HMI within the SCADA network. These steps are mostly sequential. This provides insight into the operation of the malware and allows defenders to identify points in the attack process better suited for deploying countermeasures. For example, "30. download 'newsfeed.xml'" may be used as an indicator of compromise for a Network Intrusion Detection System (IDS) that will alert if a request is seen leaving the network. Similarly, if a host-based IDS is used, the execution of CimCMSafeg.exe, "37. execute 'CimCMSafegs.exe'" may be used to identify the presence of the BlackEnergy malware. 

Additionally, due to BlackEnergy relying on manual control, there is a delay in actions due to the human in the loop. Again, the CIM representation highlights this over the original SAND models. For example, the manual step "17. Detection Prevention(S)", which is part of the automatic "15. Payload" branch, will hold up "21. Network Enumeration" and "25. Data collection" which exfiltrates data out of the network. With this insight, we can determine a weakness within the design of BlackEnergy.  

SecMoF also allows the numbering of steps which is not available in SAND or other incident response playbook representation. The analysis above demonstrates the convenience of referencing a step by its number as it, a) helps the reader to understand the order to begin investigating the diagram and b) allows each branch or node to be uniquely referred to.

By placing static images of the CIM models in a paper we lose the dynamic nature of SecMoF. For example, the full BlackEnergy image could have been reduced by placing the child nodes within the parent node, and then using SecMoF to represent the child steps as another view in eclipse. 

\subsubsection{Attack on Ukrainian Power Distribution System}\label{ukraine2015}

The Ukraine 2015 intrusion was an engineered cyber-attack into a cyber-physical system, intending to interrupt power distribution, performed by a highly skilled team with the financial backing to allow an in-depth reconnaissance of the target. The intrusion resulted in remote substations' serial-to-Ethernet converters being disabled, and the the HMI systems being erased and shut down. This was coupled with a telephone denial of service to disrupt communications. This entry point was via an email spear phishing attack, which allowed BlackEnergy to gain a foothold within the corporate and SCADA networks. 

\begin{figure*}[t]
    \includegraphics[width=\textwidth]{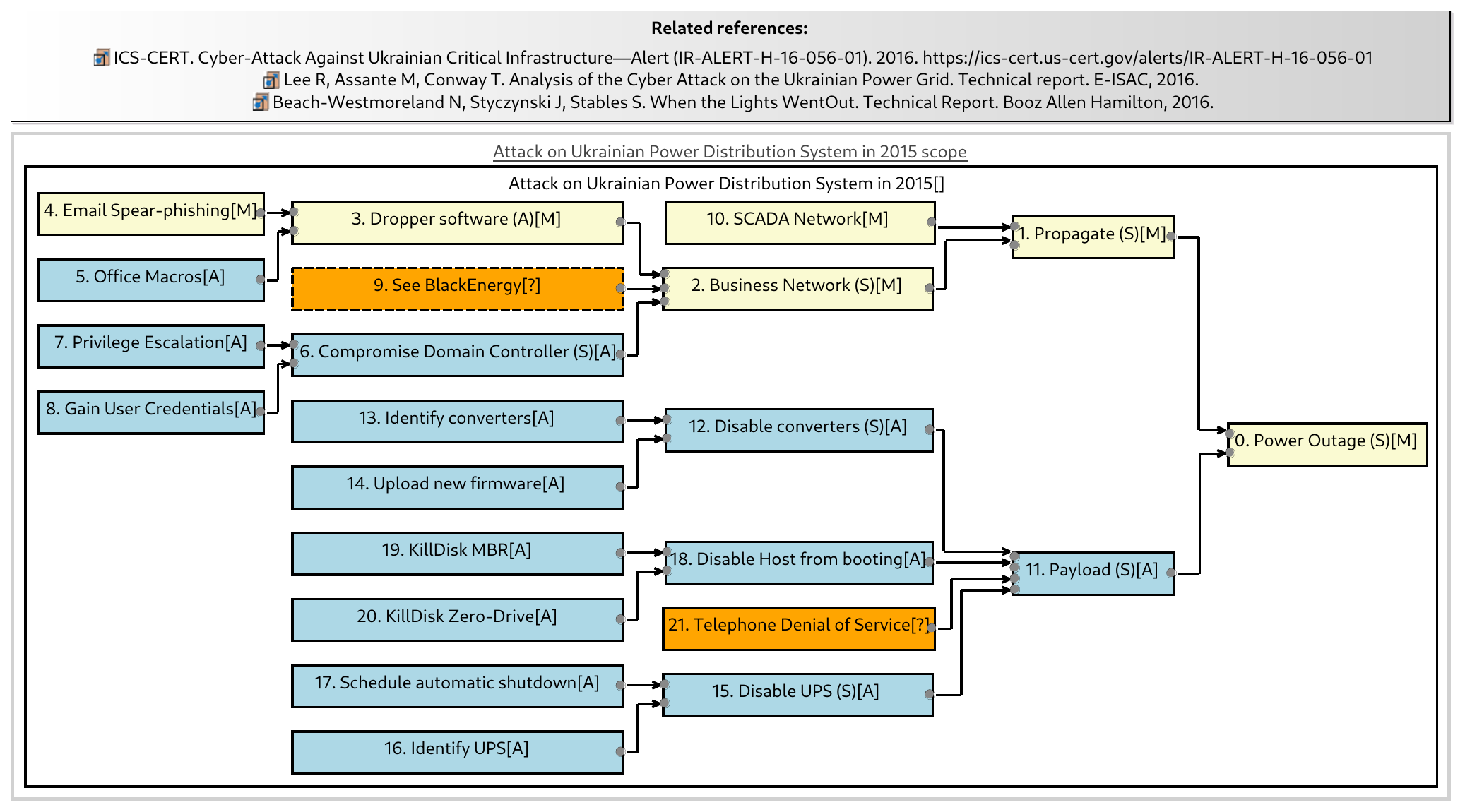}
    \caption{Attack on Ukrainian Power Distribution System in 2015 Modelled using CIM.}
    \label{fig:ukraine-fripp}
\end{figure*}

Figure~\ref{fig:ukraine-fripp} shows the CIM representation of the attack which led to the 2015 Ukrainian power outage. The attack steps were derived from the related documents listed in the "Related references" section at the top of the diagram. With the CIM model itself being converted from the SAND Attack Tree model in \cite{maynard_2020} (Figure 16). Each SAND node is represented as a CIM step, with directed arrows notating their sequence and relationships within the model. The root node is situated at the far right of the diagram, indicating the successful execution of the attack.

There are 22 steps in the process: 6 are Manual steps (yellow), 14 are Automatic steps (blue) and 2 are denoted as Unknown (orange). There is a link to a model of contributing intrusion (step 9 in Figure~\ref{fig:ukraine-fripp}) - the BlackEnergy malware attack described in Section \ref{BlackEnergy}. 

The initial impression from this CIM model is the branch of manual processes which ends with the "4. Email Spear-phishing" node. This shows that this intrusion is dependent on an attacker manually deploying "3. Dropper software" on "2. Business Network" allowing it to "1. Propagate" within both SCADA and corporate networks. The "4. Email Spear-phishing" node is a manual action from a victim, and once they click a malicious link, the remaining steps are automated, with a possible exception of the "21. Telephone Denial of Service" step. The latter step could be Manual or Automated and always originates externally to a compromised network. It was possible to quickly gain this insight due to the graphical representation, by using colouring to highlight the differing actuator types. Both node colouring and actuator types are not possible using other threat modelling such as Attack Trees. 

There were three main payloads delivered during these intrusions. First, the attackers bricked the serial-to-Ethernet converters "12. Disable converters", isolating the substations from any form of communications over Ethernet. Second, the HMI servers were disabled "18. Disable Host from booting", erasing their master boot records. Third, the backup power supply disabled "15. Disable UPS" ensuring that when the attackers shut down the substations, via access to the HMI, the HMI and other systems will also go offline. 

While this attack resulted in a disruption to the Ukraine power grid, the actions performed were not sophisticated in terms of technical ability. BlackEnergy gave them access to the HMI through a vulnerability in the service, afterwards, the steps performed could have been applied to any IT system. 

\subsection{Impact of Attacks and Dependency Model Analysis}\label{s:dependency-model-analysis}
We highlight the impact of the intrusions using the configurable dependency model of SCADA systems presented in \cite{cherdantseva_2022} that was developed by collecting and analysing the understanding of the dependencies within a SCADA system from 36 domain experts. The dependencies were identified through workshops with experts. The SCADA dependency model has 452 elements covering the six key areas of a SCADA system: 

\begin{enumerate}
  \item Management
  \item Data (Information)
  \item System Life Cycle 
  \item Employees
  \item External Dependencies (Environment)
  \item System Architecture
  \begin{itemize}
    \item Networks
    \item Software
    \item Hardware
  \end{itemize}
\end{enumerate}

The attack impact analysis in the following sub-sections focuses on the impact only on the Security characteristics within the Network component of the System Architecture subset of the model because this is sufficient to demonstrate the advantages of linking a CIM with a dependency model.

The impact is represented as a change in the state of a Paragon. The default value of a Paragon is success, meaning its state is '1' (green). An impact on the dependency model can be completely failed '0' (red), or a percentage of success, e.g. 0.1 to 0.9. 

\subsubsection{BlackEnergy Malware Attack}\label{BlackEnergy_dm}
\begin{figure*}[t]
    \includegraphics[width=\textwidth]{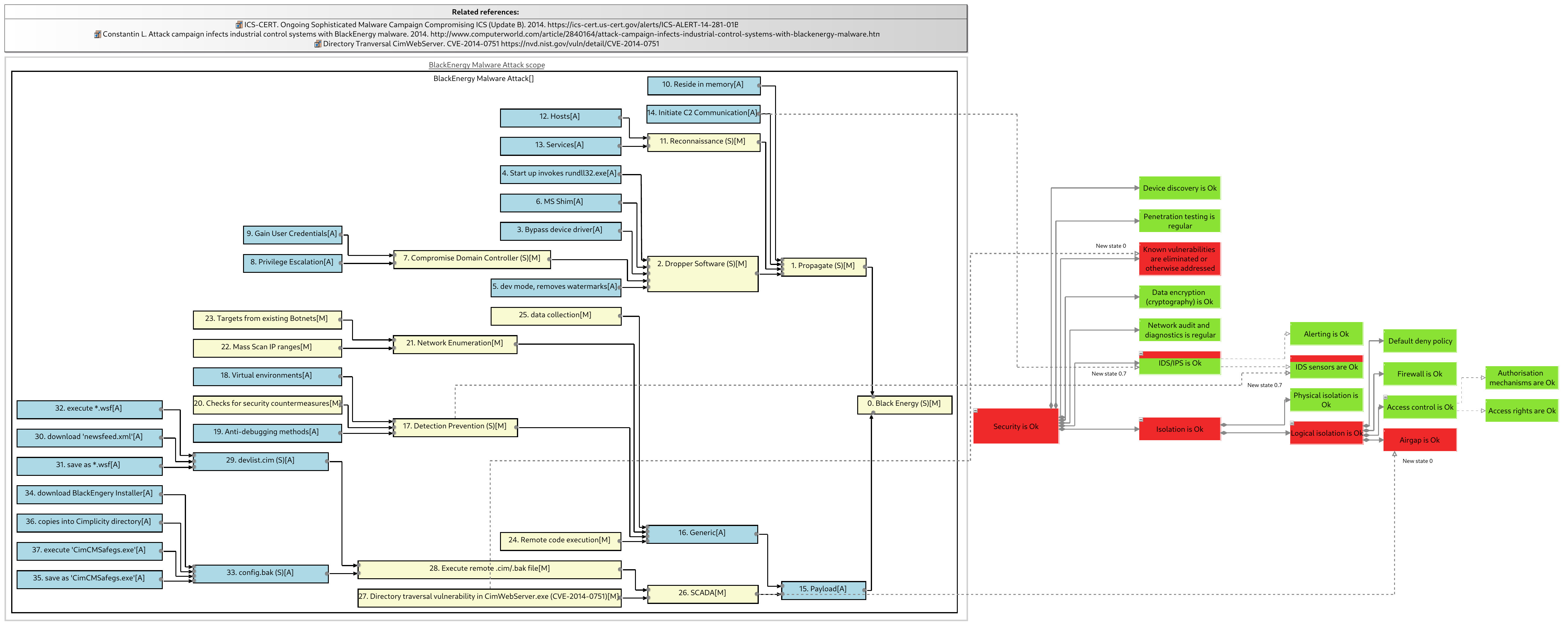}
    \caption{BlackEnergy attack CIM and its impact on the 'System-Architecture' dependency model.}
    \label{fig:blackenergy-dm}
\end{figure*}

BlackEnergy Malware is a flexible malware with support for exploiting a vulnerability within a SCADA HMI. Figure~\ref{fig:blackenergy-dm} shows (1) the BlackEnergy attack process model captured in the CIM format on the left; (2) the \texttt{System Architecture/Network/Security} dependency model on the right, and (3) and the mapping between the two indicating the impact of particular steps of the attack on the state of affected dependencies. 

The first impact on the dependency model is by the "1. Propagate" branch of the CIM when the adversary "14. Initiates a C2 Communication channel". This impacts the state of the dependency model node \texttt{IDS/IPS is Ok} from probability 1 to probability 2 because the adversary can initiate communication without being detected or prevented by the existing countermeasures which reduces the probability of this node being in a successful state (when the probability of success equals 1). 
This allows the adversary to further manually embed themselves into the system by performing the "Detection Prevention" step (node 17) along with its child steps (nodes 18-20), resulting in additional data being exfiltrated into, and out of the network. This action being undetected negatively impacts the \texttt{IDS sensors are Ok} Paragon.

Now that the adversary has gained remote access to the network this nullifies the \texttt{Isolation is Ok/Logical isolation is Ok/Airgap is Ok} Paragon, meaning that there is no logical separation between the attacker and their target as shown by node "26. SCADA" (meaning, the attacker has direct access to the SCADA system). 
Finally, the attackers exploit a vulnerability – as indicated by node 27 "Directory traversal vulnerability in CimWebServer.exe", resulting in \texttt{Known vulnerabilities are eliminated or otherwise addressed} becoming invalidated.

The BlackEnergy CIM analysis impacts only the networking security Paragons of the Dependency Model. Then, in turn, the impact is propagated up to the root Paragon of a \texttt{Secure and safe operation of a SCADA system}. This provides ample opportunities for countermeasures to be considered either during deployment or at the development phase.

\subsubsection{Attack on Ukrainian Power Distribution System}\label{ukraine_dm}

\begin{figure*}[!h]
    \includegraphics[width=\textwidth]{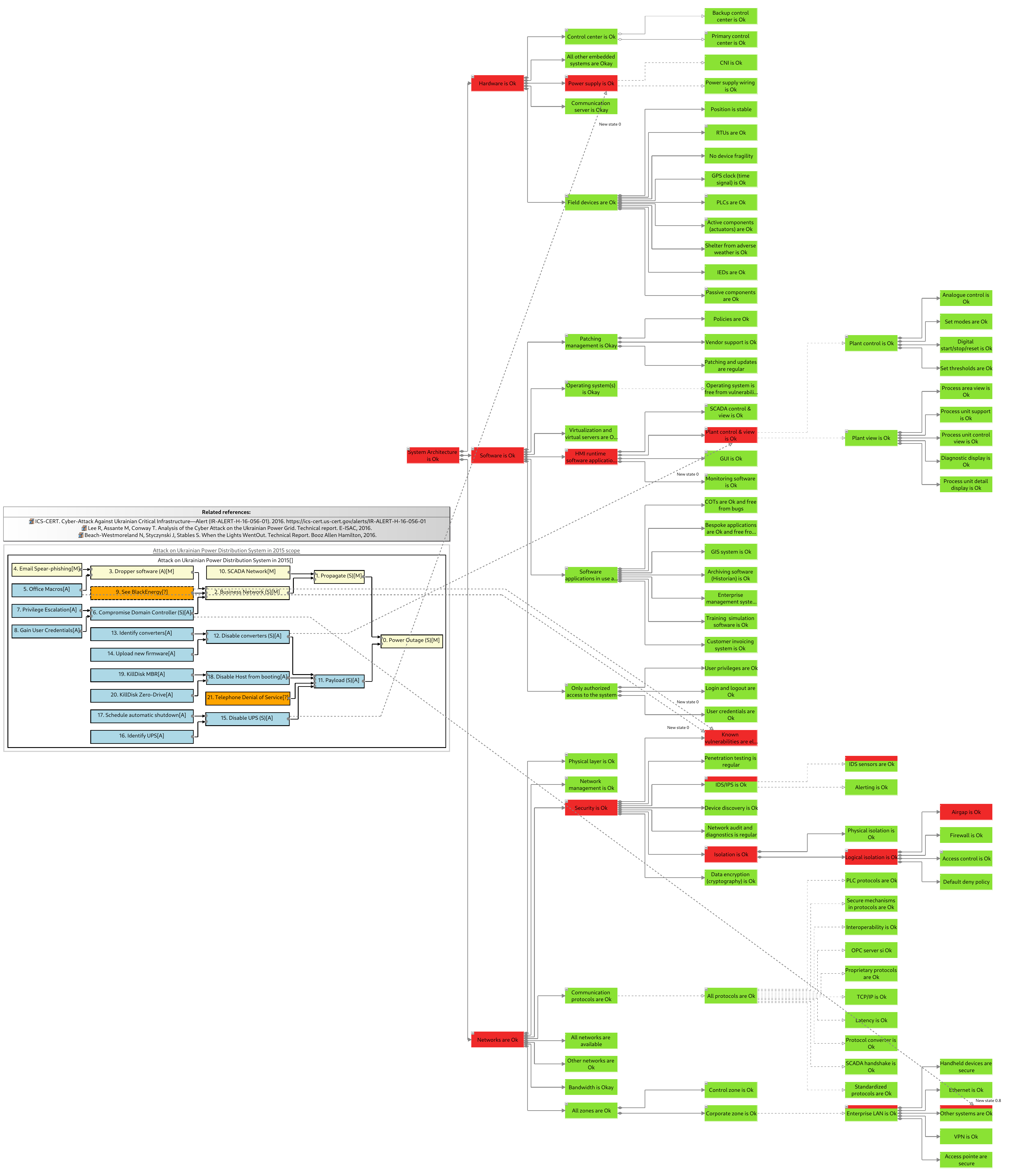}
    \caption{Ukrainian attack CIM and its impact on the 'System-Architecture' dependency model.}
    \label{fig:ukraine-dm}
\end{figure*}

The Ukraine 2015 power outage intrusion is mapped to the Dependency Model in Figure~\ref{fig:ukraine-dm}. As this was an intrusion which resulted in the loss of power, we can map the root CIM node to the root Dependency Model \texttt{Secure and safe operation of a SCADA system} as the SCADA system is not secured. We have hidden many of the SCADA dependency model nodes which were not affected by this attack to make it easier to display in this paper.
As this intrusion leverages the BlackEnergy malware, some of the nodes within the Dependency Model have been marked as failures, but do not have a direct mapping to this CIM. Consider Steps 9 and 3 as already mapped within the BlackEnergy analysis. This is reflected in the Dependency Model e.g. \texttt{Security is Ok} and \texttt{Airgap is Ok}.

Step 6 of the intrusion compromises the business network's domain controller. However, in the current SCADA dependency model, there is no appropriate node for this step to be directly mapped to. The closest available node is the \texttt{Other systems are Ok} Paragon under the branch of \texttt{Networks are Ok/All zones are Ok/Corporate zone is Ok/Enterprise LAN is Ok}, which was used instead. As we are modelling the compromise of just the domain controller we set the impact to the Paragon as 0.8, from 1.  

The attackers disabled the serial to Ethernet converters, at step 12. This prevents the HMI from viewing the latest data regarding the plant. Accordingly, this is mapped to the \texttt{Software is Ok/HMI runtime software application(s) is Ok/Plant control \& view is Ok} Paragon.
Finally, step 15 disables the backup power system resulting in the \texttt{Hardware is Ok/Power supply is Ok} failure.

By mapping the two intrusions together on the same Dependency Model we see how an adversary can chain weaknesses in different parts of a system to negatively impact the operation of a distributed system, such as a SCADA network. 

\section{Discussion and Lessons Learned}\label{s:discussion}

In this section review the lessons learned from modelling SAND Attack Trees using SecMoF to model the intrusions using CIM and map them to a dependency model. We have organised the discussion into three sub-sections: 
\begin{itemize}
  \item \textbf{Conceptual insights}: Focus on the methodology and underlying principles; 
  \item \textbf{Technical insights}: Focus on the tooling and data formats.
  \item \textbf{Overview insights}: Focus on the remaining seven intrusions.
\end{itemize}

\subsection{Conceptual Insights}

Modelling SAND Attack Tress (SAND) using CIM provides conceptual insights concerning the potential role of attack modelling and how it can be improved in support of a holistic security framework. Table~\ref{tab:concept} provides a concise overview of features that we consider essential for effective attack modelling. The method column indicates whether a specific feature is supported by CIM, SAND, or Both of the methods. In the remainder of this section, we explain each of these points in greater detail.

\begin{table}[t]
\caption{Summary review regarding the concepts.}
\begin{tabular}{p{0.2cm}p{10cm}l}
\toprule
\textbf{ID} & \textbf{Summary} & \textbf{Method} \\
\midrule
C1 & Able to align threat actions to a risk assessment model. & CIM \\
C2 & Threat modelling actions can directly contribute to the development of IR playbooks. & CIM \\
C3 & Ordering of attack steps. & Both \\
C4 & Ability to add Manual, Automatic, and Unknown actuators. & CIM \\ 
C5 & Informative graphical representation. & Both \\
C6 & Lack of mechanism to reference whole or a subset of existing models. & Both \\
\bottomrule
\end{tabular}
\label{tab:concept}
\end{table}

One of the primary drivers for this work is (C1) the ability to align threat models along with the broader risk assessment without losing the nuances of the threats. This can improve the quality of the risk assessment and the alignment of the threat with the specifics of the system under assessment. This becomes important in the CNI context due to the nature of specific systems being unique in many ways. By gaining an understanding of an intrusion in this way, operators can better determine the impact that such an intrusion may have on their system.

Likewise, incorporating intrusions from threat modelling analysis (C2) into IR playbooks ensures that learning is not lost by using the same approach for both threat modelling and IR playbooks. This allows a SOC Analyst to quickly refer to other known intrusions while creating new playbooks. 

Both of the methods allowed the ordering of the attacker steps (C3). This is the defining feature of SAND in that it defines the sequential AND relationship, which allows the expression of more complex attack trees as opposed to the original AND, and OR relationships.  

Unlike CIM, SAND does not have an option to assign an actuator to a node (C4). This results in less information being stored in the attack model itself. With SAND, this type of information would be included along with associated documentation. Having this information captured in a model like CIM allows for quicker identification of an attacker's path or objective. When importing the SAND models into CIM additional research was performed to identify if a SAND node was manual or automatic. If not enough information was found, it was marked as unknown. While not used much in the examples, it is important to note that more than one property may be assigned to each node. This allows for the modelling of more complex intrusions than may be possible with SAND alone.

Both methods represent the attack steps in a graphical presentation, with SecMoF CIM can colour code the actuator types resulting in a more visual indication of the attack paths (C5). It is interesting to note that when the intrusions were assigned actuators, typically the parent would be manual and the children automated. This depends on the type of intrusion, e.g. Spear-phishing requires the victim to click on the email before the automated malware can be deployed. The advantage of this is that the models now provide additional insight into how a system could be protected, as this manual action could be used to improve countermeasures. This feature could be enhanced by adding another symbol to indicate whether it is the attacker or the victim performing the step.

Each of the nine intrusions share a common objective, at least in part due to their victims. This shows in some of the intrusion models as steps that are linked to other models. Typically, models which refer to others will state it, and expect the reader to navigate and review the model themselves. If SAND and CIM were able to reference existing models, either wholly or as a subset, this would reduce the repeated steps while allowing the rendering of more concise models (C6).

\subsection{Technical Insights}

In this section, we review the lessons learned from converting tab-indented attack trees using the Security Model Converter (SMC) and editing them using the Security Modelling Framework (SecMoF) tool. Table~\ref{tab:tech} provides a concise overview of the technical review, showing a summary of each point, which tool relates to SMC, SecMoF, or Both. 

\begin{table}[t]
\caption{Summary review regarding the technical approach.}
\begin{tabular}{p{0.2cm}p{10cm}l}
\toprule
\textbf{ID} & \textbf{Summary} & \textbf{Tool} \\
\midrule
T1& Nodes should be numbered for non-interactive viewing. & SecMoF \\
T2& Expand supported associated reference types. & SecMoF \\
T3& Placement of the SAND Operators are counter-intuitive. & Both \\
T4& Readability of SAND models are lost due to node placement. & SecMoF \\
T5& Associated references are truncated when exported to an image. & SecMoF \\
T6& Missing and extra artefacts left over from conversion. & SMC \\
\bottomrule
\end{tabular}
\label{tab:tech}
\end{table}

By design, the SecMoF can verify if a model is correct using its meta-model. However, when a model is viewed statically (e.g. when included within a paper), this information is lost to the reader. In most cases, intrusion models are viewed statically within an associated document. (T1) So for this case each node being numbered is desired as it allows easier explanation of the attack.

The SAND operator is represented at the end of the node's description, as SecMoF does not take into account dependencies between activities, including sequential AND relationships. (T3) This is counter-intuitive as the reader will view the model and quickly identify the relationships and the actions, before understanding the ordering constraints. This overhead is further amplified by the node placement disregard of the sequential ordering, which is indicated in visual SAND Attack Tree representations (T4).  

Currently, each reference used within a model is specified as a free-form text. Allowing well-structured references (T2, T5) to be included in the models would benefit the user. Specifically, capturing the URL, Title, Publisher, and DOI could help the user locate additional documentation reliably.

When the SAND Attack Trees were converted by SMC (for this article), there were several additional nodes created which needed to be removed (T6). Accordingly, the node numbering was incorrect due to SMC incrementing the node number value for every node, including those removed from the original SAND text model. These issues have been addressed and patched in the current SMC version. 

\subsection{Overview Insights}\label{overview-insight}
 
For this paper, we converted nine SAND Attack Trees into CIM models and analysed them in SecMoF. Due to limited space, we discussed in detail two BlackEnergy and the 2015 Ukraine power outage. The remaining seven attack CIM models are shown in Appendix~\ref{appendix:remain}. They are Stuxnet, Duqu2.0, German Steel Mill, Havex, CrashOverride, TRISIS, and the second Ukrainian power outage. This section will briefly highlight some of the insights from these models.

Stuxnet is one of the most well-known intrusions and as such has plenty of literature analysing it. Because of this, and due to its complexity, the SAND models were originally split into two separate trees. Part A focused on the business network, and Part B the OT network. This highlights the need for SAND and CIM to support linking to existing models. Stuxnet was designed to operate autonomously with minimal manual interactions, many of the steps are automatic, while the steps that are manual such as User Inserts USB or User opens file project are manual actions performed by the victim. 
Havex was a small watering hole and email spear-phishing intrusion that targeted industrial networks. Similar to Stuxnet, Havex shows that it would be beneficial for CIM to include actuators such as Manual-Attacker and  Manual-Victim.

Duqu2.0 is a sophisticated cyberespionage toolkit similar to BlackEnergy. The CIM model is more feature-complete than BlackEngery as it shows, at a high level, the features it supports. As it was designed to work without detection, many of the features are triggered manually by the attacker with the lower-level details being automated by the malware itself. 
As with the second Ukraine power outage in 2017, many of the steps had been replaced with automated steps, except monitoring the IT staff behaviours before performing actions as they were able to blend in with their actions, as at the time there was an incident response operation in progress. The second intrusion follows a similar method as the first. 

Having converted these attack trees into CIM models we can further identify commonalities between each intrusion for use in developing a more robust risk assessment and incident response playbooks. 

\section{Conclusion and Future Work}\label{s:conclusion}

We proposed a novel way to represent SAND Attack Trees (SAND) using the FRIPP meta-model and Operations-Informed Incident Response (OIIRP) tooling. In doing so we resolve the disconnect between IR playbooks and threat analysis that are traditionally performed using distinct methods and at various stages of the systems' lifecycle. By using Compatible Intrusion Models (CIM, a variation of FRIPP) to model Attack Trees, we can embed additional information within the models as well as enhance the original analysis to pinpoint where an adversary may be required to perform manual actions or where they may be able to automate actions. Together this unifies the three main cyber security categories: attack modelling, incident response, and risk assessment. This advances the state-of-the-art with respect to the communication gap established in a recent systematic literature review \cite{khalil_2024}.

We have identified points of potential improvements to both CIM and SAND in terms of improving the representation of the models for capturing additional data. Additionally, we advance the state of tooling for CIM by identifying User Experience (UX) issues. Based on this work, we continue the development of SecMoF and SMC to improve the resulting models and UX of both IR Playbooks and threat models created using OIIRP. 

Finally, we created an initial public knowledge base of CIM intrusion models along with open-source tooling for converting tab-indented Attack Trees into CIM models. This knowledge base may be used for further research into both Incident Response (IR) and threat modelling within the critical national infrastructure industry. 

\section{Acknowledgements}
This work was supported by the Engineering and Physical Sciences Research Council (EPSRC) (Grant number EP/V038710/1).
\bibliography{intrusion-modelling.bib}

\section{Appendix: Additional Converted Models}
\label{appendix:remain}
The full set of converted model is presented below. The colour coding for 'manual' and 'automatic' tasks is not included on representations below.

\begin{figure*}[t]
    \includegraphics[width=\textwidth]{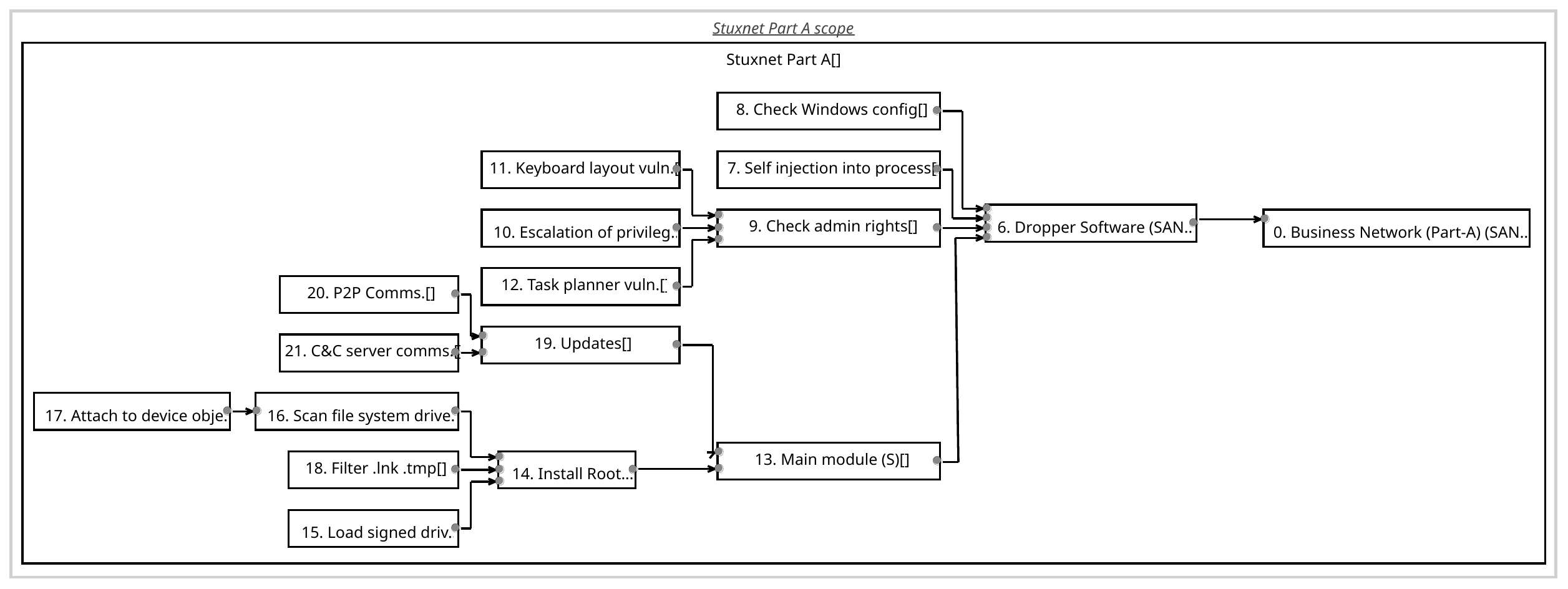}
    \caption{Stuxnet (Part-A) Business Network}
\label{fig:Busi}
\end{figure*}

\begin{figure*}[t]
    \includegraphics[width=\textwidth]{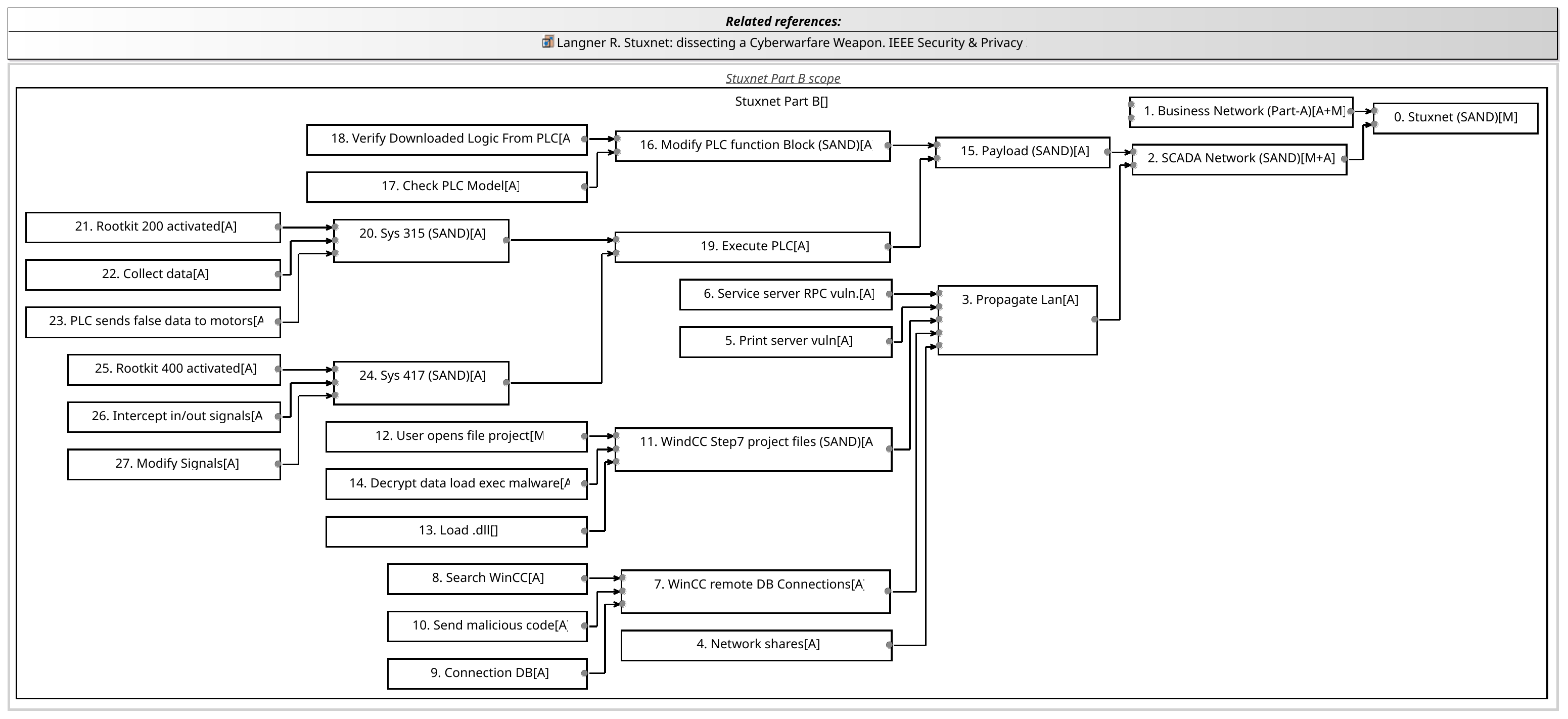}
    \caption{Stuxnet (Part B) OT Network}
    \label{fig:Stux}
\end{figure*}

\begin{figure*}[t]
    \includegraphics[width=\textwidth]{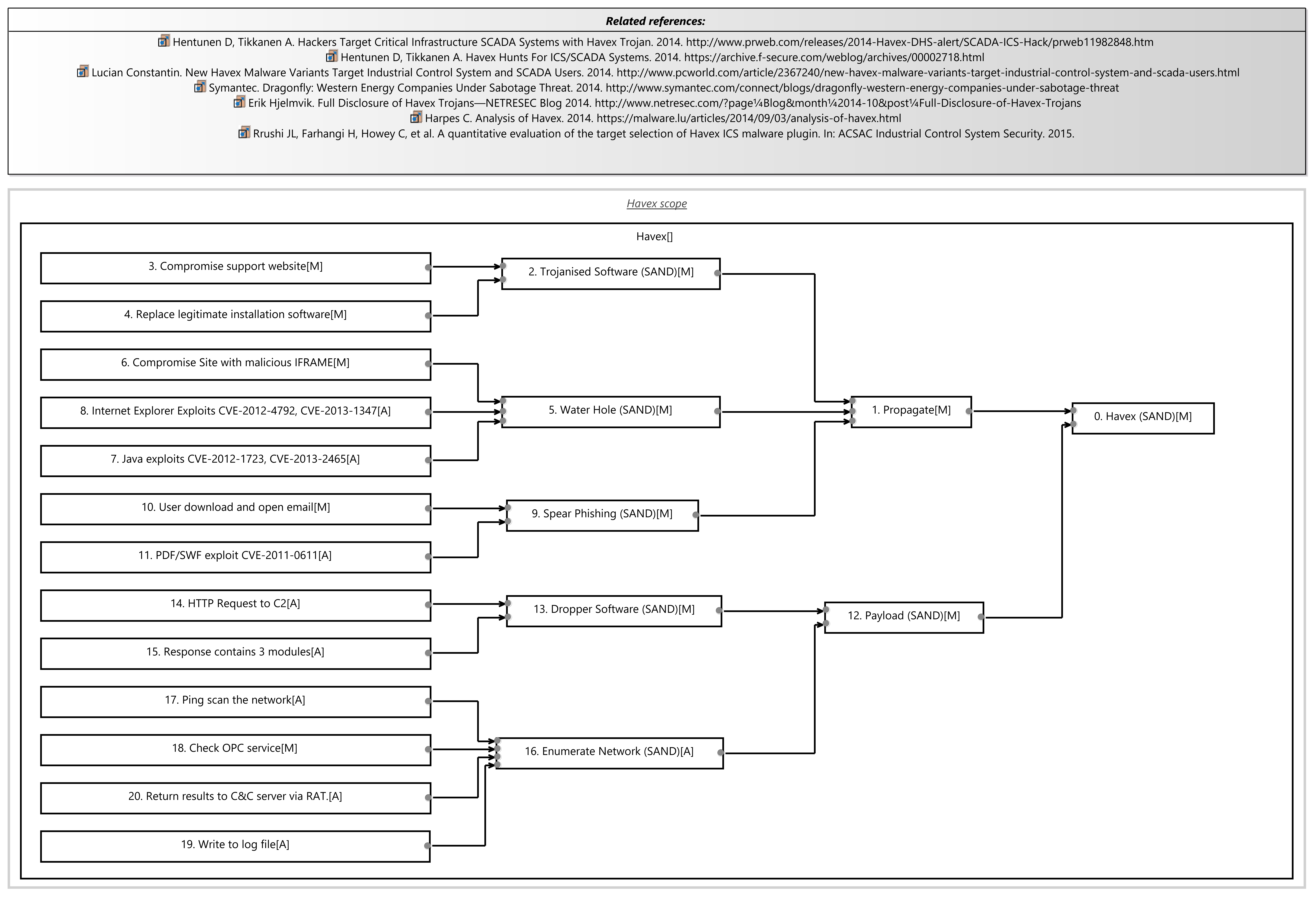}
    \caption{Havex}
    \label{fig:Have}
\end{figure*}

\begin{figure*}[t]
    \includegraphics[width=\textwidth]{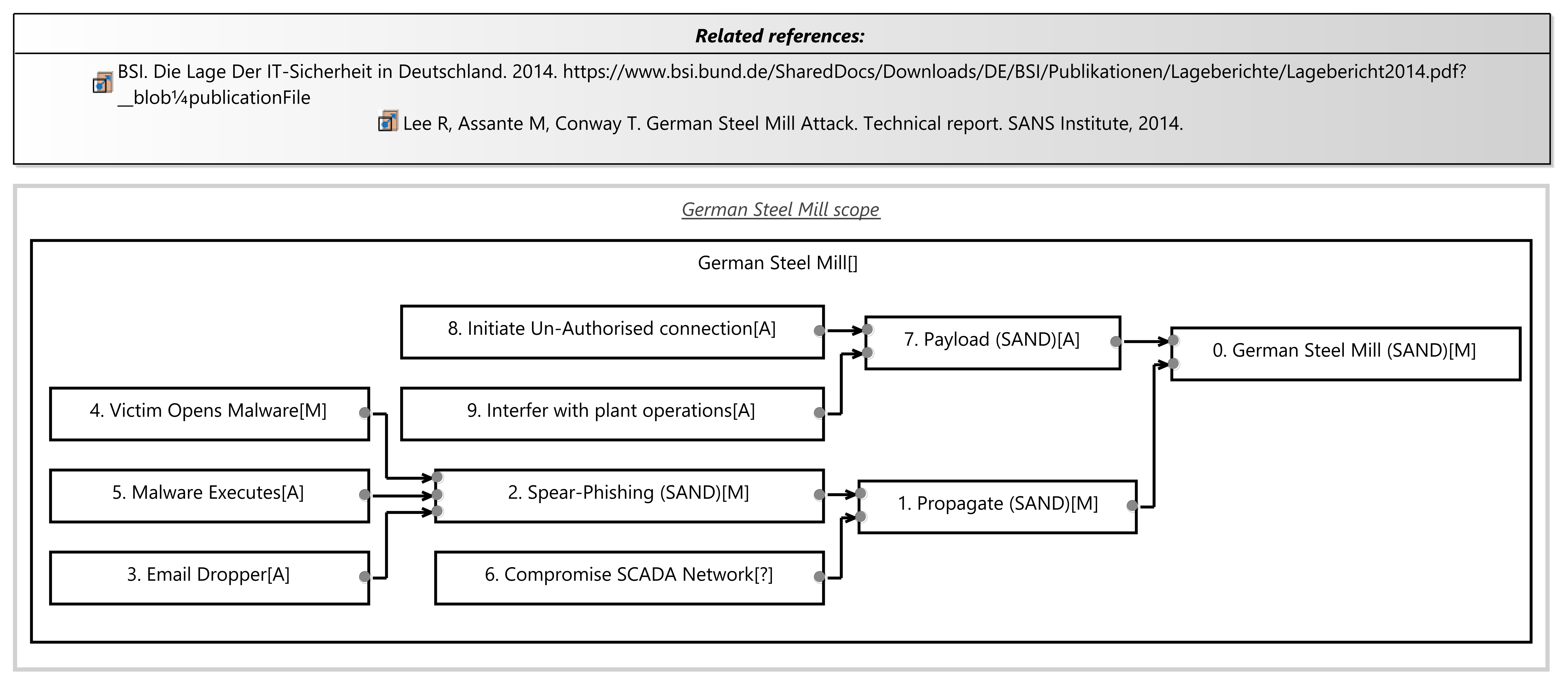}
    \caption{German Steel Mill}
    \label{fig:germ}
\end{figure*}

\begin{figure*}[t]
    \includegraphics[width=\textwidth]{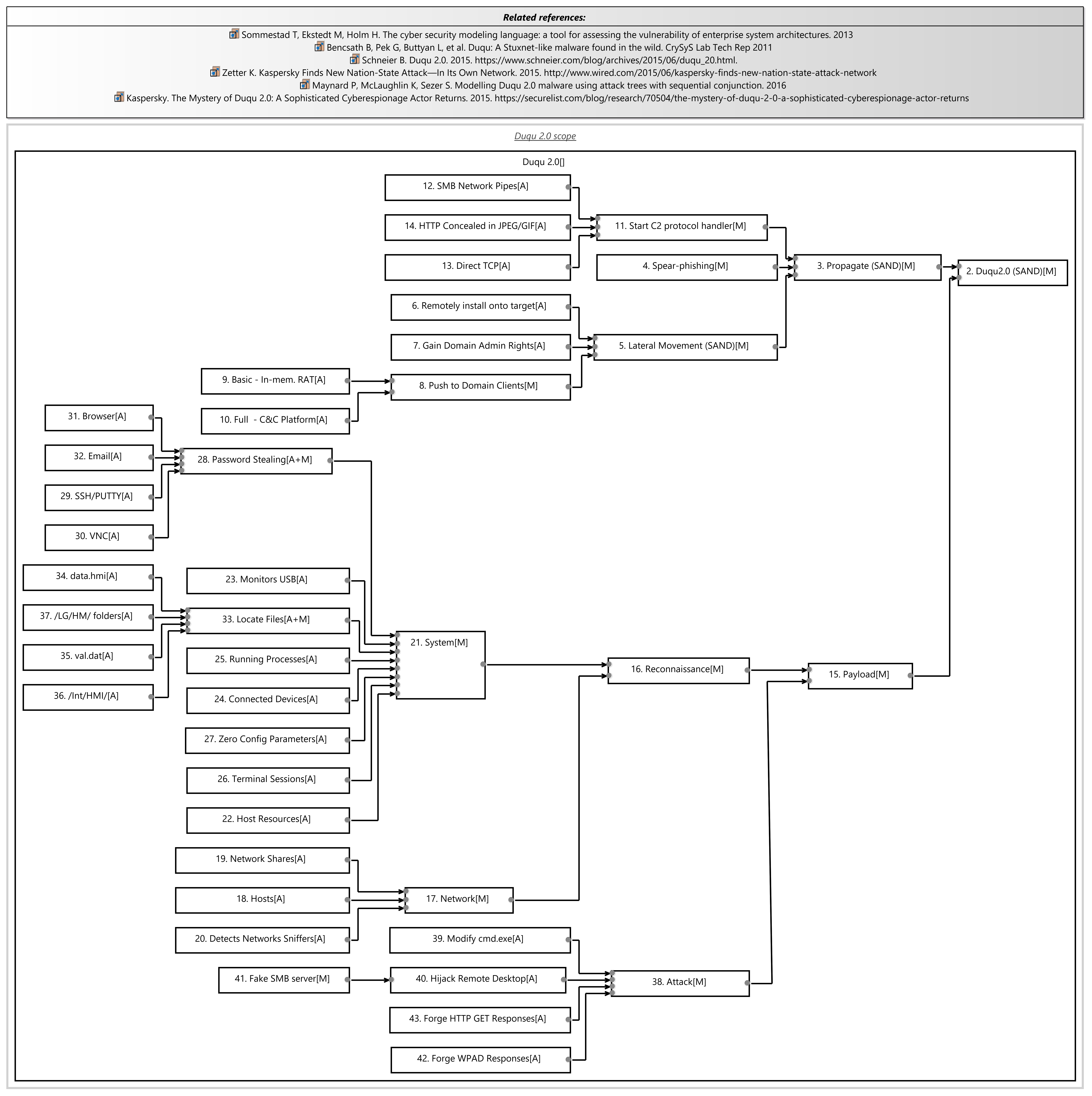}
    \caption{Duqu2.0}
    \label{fig:duqu}
\end{figure*}

\begin{figure*}[t]
    \includegraphics[width=\textwidth]{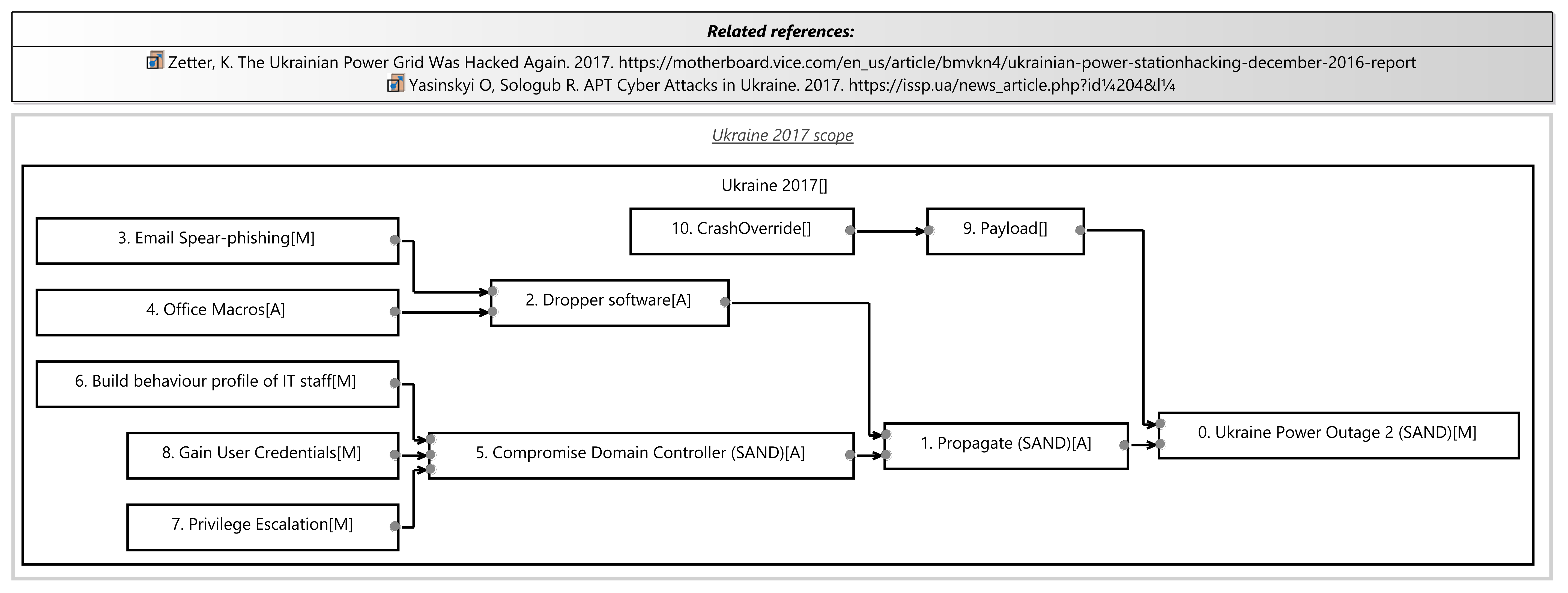}
    \caption{Ukraine 2}
    \label{fig:ukra}
\end{figure*}

\begin{figure*}[t]
    \includegraphics[width=\textwidth]{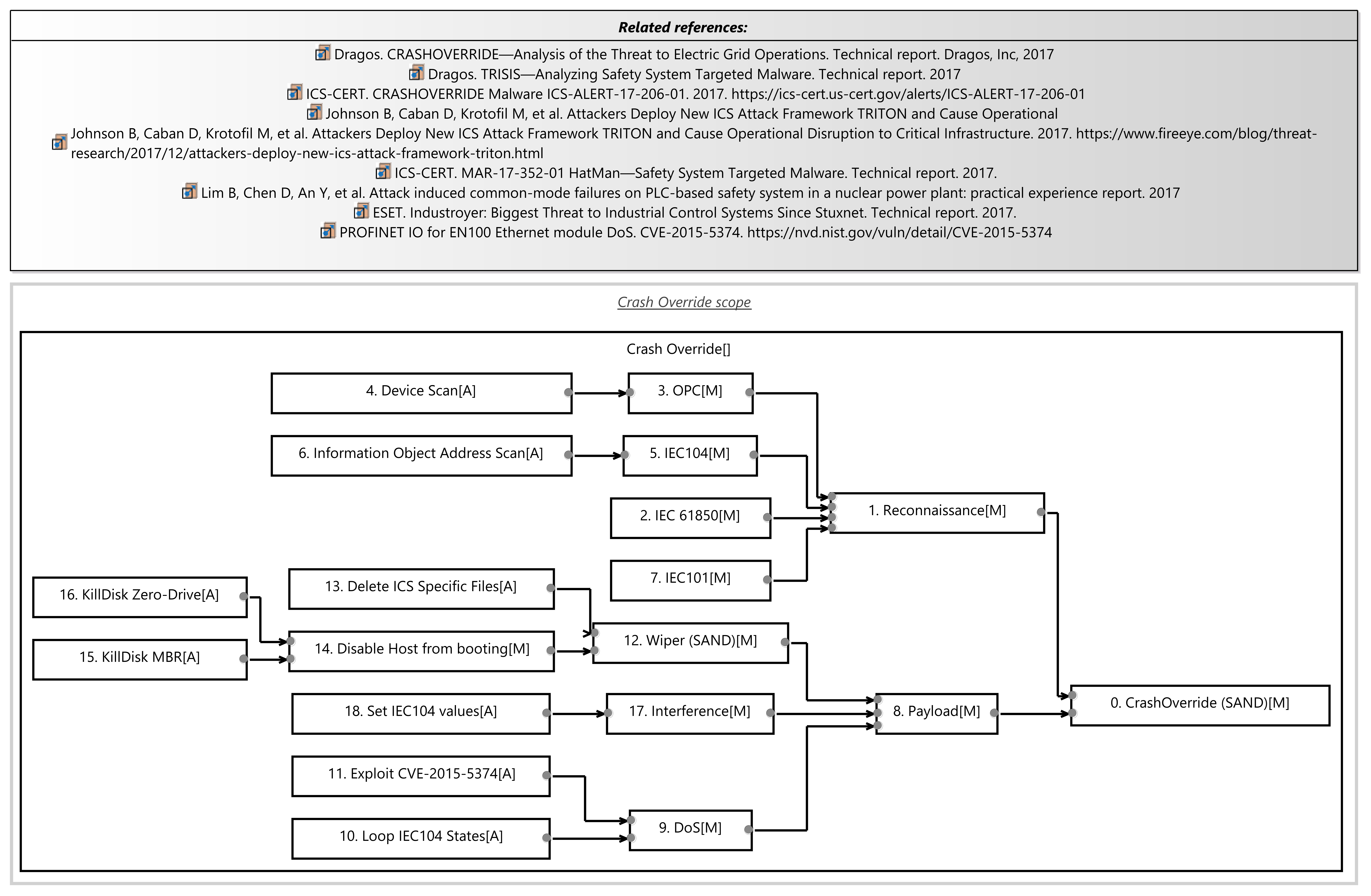}
    \caption{CrashOverride}
    \label{fig:imgc}
\end{figure*}

\begin{figure*}[t]
    \includegraphics[width=\textwidth]{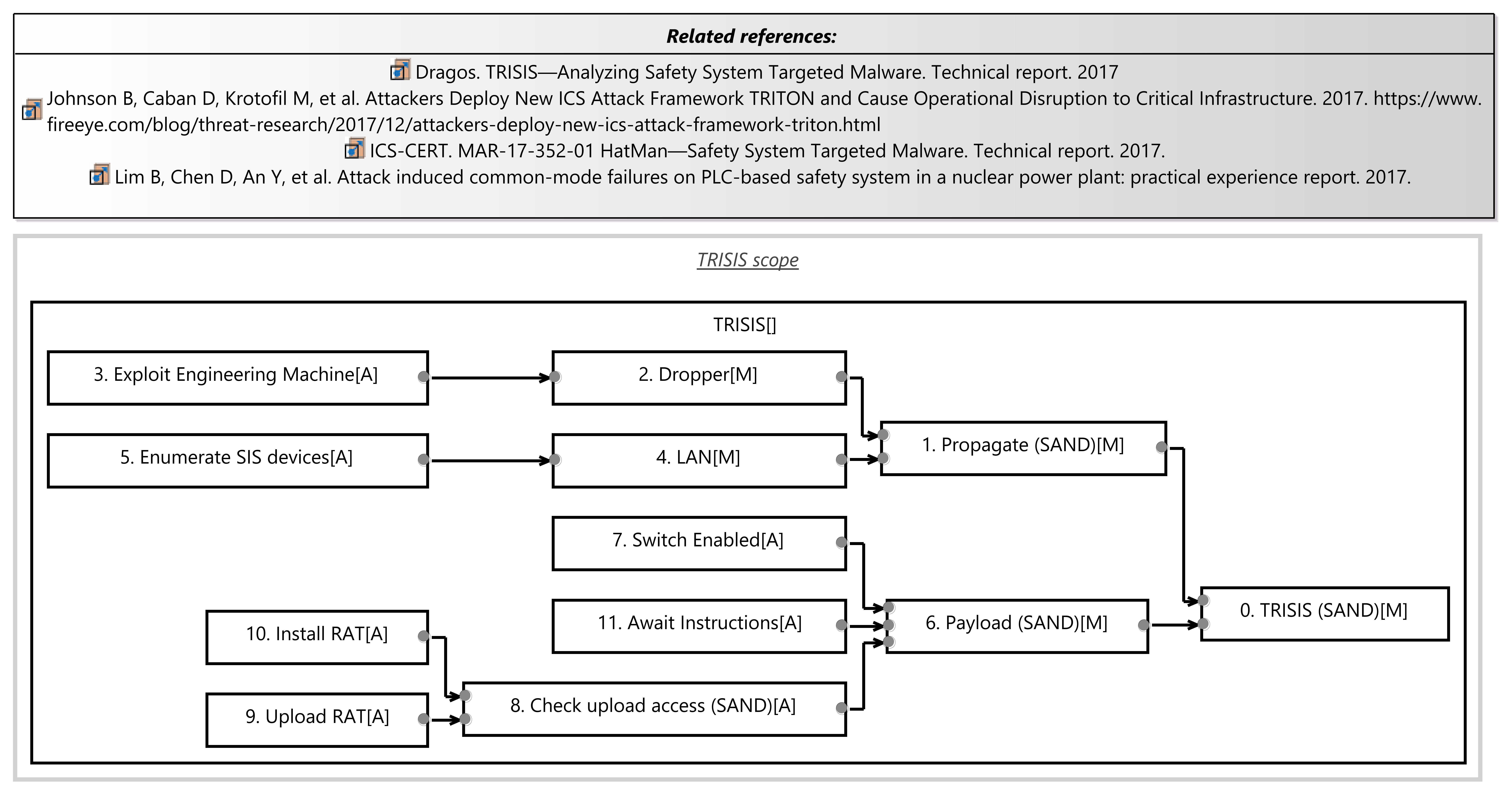}
    \caption{TRISIS}
    \label{fig:tris}
\end{figure*}

\end{document}